\documentclass[fleqn,usenatbib]{mnras}
\usepackage{newtxtext,newtxmath}
\usepackage[T1]{fontenc}
\usepackage{ae,aecompl}
\usepackage{rotating}
\usepackage{graphicx}	
\usepackage{amsmath}	
\usepackage{amssymb}	
\usepackage[english]{babel}

\title[Sticky or not sticky? Measurements of the tensile strength of micro-granular organic materials]{Sticky or not sticky? Measurements of the tensile strength of micro-granular organic materials}

\author[D. Bischoff et al.]{
D. Bischoff$^{1}$\thanks{E-mail: d.bischoff@tu-bs.de},
C. Kreuzig$^{1}$,
D. Haack$^{2}$,
B. Gundlach$^{1}$,
J. Blum$^{1}$\\
$^{1}$ Institut f\"ur Geophysik und extraterrestrische Physik, Technische Universit\"at Braunschweig, Mendelssohnstr.3, 38106 Braunschweig, Germany\\
$^{2}$ Deutsches Zentrum f\"ur Luft- und Raumfahrt, Rutherfordstra\ss e 2, 12489 Berlin-Adlershof, Germany
}

\date{???}

\pubyear{2020}
\begin{document}
\label{firstpage}
\pagerange{\pageref{firstpage}--\pageref{lastpage}}
\maketitle

\begin{abstract}
Knowledge of the mechanical properties of protoplanetary and cometary matter is of key importance to better understand the activity of comets and the early stages of planet formation. The tensile strength determines the required pressure to lift off grains, pebbles and agglomerates from the cometary surface and also describes how much strain a macroscopic body can withstand before material failure occurs. As organic materials are ubiquitous in space, they could have played an important role during the planet formation process. This work provides new data on the tensile strength of five different micro-granular organic materials, namely, humic acid, paraffin, brown coal, charcoal and graphite. These materials are investigated by the so-called Brazilian Disc Test and the resulting tensile strength values are normalised to a standard grain size and volume filling factor. We find that the tensile strength of these materials ranges over four orders of magnitude. Graphite and paraffin possess tensile strengths much higher than silica, whereas coals have very low tensile strength values. This work demonstrates that organic materials are not generally stickier than silicates, or water ice, as often believed.

\end{abstract}


\section{Introduction}\label{Sect_1} 
Organic matter is defined as chemical compounds containing carbon-hydrogen bonds. On Earth, organic materials are carbon-based compounds that have come from the remains of organisms and their waste products. In space, organic compounds are ubiquitous, as they are found in diffuse clouds, in the envelopes of evolved stars, in dense star-forming regions, in protoplanetary discs, in comets, on the surfaces of minor planets, in meteorites and in interplanetary dust particles \citep{vanDishoeck2008}. Stars at different evolutionary phases are able to produce complex organic compounds and eject them into space, filling the regions between the stars. The compounds are so complex that their chemical structures can even resemble the makeup of coal and petroleum. These materials are then incorporated into star- and planet-forming regions and, hence, also into solid objects, such as comets, asteroids and planets. 
\par
Besides the discussion about the origin of life on Earth, organic matter is also often believed to be stickier than other solid compounds, such as siliceous materials, so that mutual collisions in protoplanetary discs may lead to larger agglomerates (see below). One method to evaluate the stickiness of materials in the laboratory is the so-called Brazilian Disc Test, which is designed to measure the tensile strength of rigid matter, such as stones, but also of compressed granular samples \citep{Meisner2012,Gundlachetal2018, Haacketal2020}. The tensile strength is an important material property, because it is defined as the maximal mechanical tension a material can resist before it breaks. Hence, the value of the tensile strength determines whether an object can withstand a given stress. This is why the tensile strength is the key material parameter to understand the size distribution of small solar system objects during their collisional evolution in debris discs \citep[see, e.g.,][]{Bottkeetal2005}. 
\par
The tensile strength also plays an important role during the planet-formation process. \citet{Kimura2015} and \citet{Homma_2019} found that an organic mantle around silica grains can lead to an increase of the fragmentation threshold velocity, shown experimentally with interstellar grains by \citet{Kouchi2002}. This could open a door for a direct growth of these particles into planetesimals. In a recent work, \citet{Kimuraetal2020} showed that the tensile strength of dust aggregates decreases with increasing volume of the object, which depends on the so-called Weibull modulus, a parameter that describes the variability of the material strength. The minimum tensile strength of the organic matter of cometary interplanetary dust particles with an aggregate diameter of $\sim 5 \,\mathrm{\mu m}$ was estimated by \citet{Flynn2013} to be $\sim 150$ to $325 \,\mathrm{Pa}$. \citet{kudo2002} found the tensile strength of organic coating on copper spheres to be $10^5 \,\mathrm{Pa}$. Data from the COSIMA instrument onboard Rosetta was recently interpreted by \citet{kimura2020b} who found a reduction of the sticking and fragmentation threshold velocity by a carbonisation of the organic matter.
\par
The tensile strength can be also understood as a measure for the stress that needs to be applied to detach two particles from each another \citep{Blumetal2006}. This is why cometary activity, namely the ejection of dust from the surface of a comet nucleus by the sublimation of volatiles can only be understood if the tensile strength of the material is considered in the calculations. Dust ejection can only occur if the local gas pressure exceeds the local tensile strength of the material \citep[see, e.g.,][]{Fulleetal2019,Fulleetal2020a,Fulleetal2020b,Gundlachetal2020}. Approximately fifty percent of the non-volatile mass of cometary nuclei is made of organic materials \citep{Bardynetal2017}. Unfortunately, the exact composition of the non-volatile, organic cometary matter is unknown. Studying carbonaceous chondrites, the most primitive and least processed meteorites, may provide the closest picture of the chemical composition of non-volatile, organic cometary matter. Most of the organics found in carbonaceous chondrites are in an insoluble macro-molecular form, often described as kerogen-like \citep{vanDishoeck2008}. The remaining part consists of soluble organics, such as corboxylic acids, PAHs, fullerenes, purines, amides and other prebiotic molecules \citep{Bottaetal2002}.
\par
This is why in the past years the question arose, whether the presence of organic compounds can lead to an increase, or decrease, of the tensile strength (stickiness) of the cometary surface material. In this paper, we present new measurements of the tensile strength of organic materials, which were carried out in the framework of the CoPhyLab project. First, the experimental procedure and the selected samples are described in Section \ref{Sect_2}. Then, the results of the experiments are presented in Section \ref{Sect_3}, which is followed by a discussion of how these findings can help to improve our understanding of the comet dust ejection process (Section \ref{Sect_4}). The paper ends with a concluding Section \ref{Sect_5}.

\section{Experimental technique}\label{Sect_2}
\begin{figure}
    \centering
    \includegraphics[width=\columnwidth]{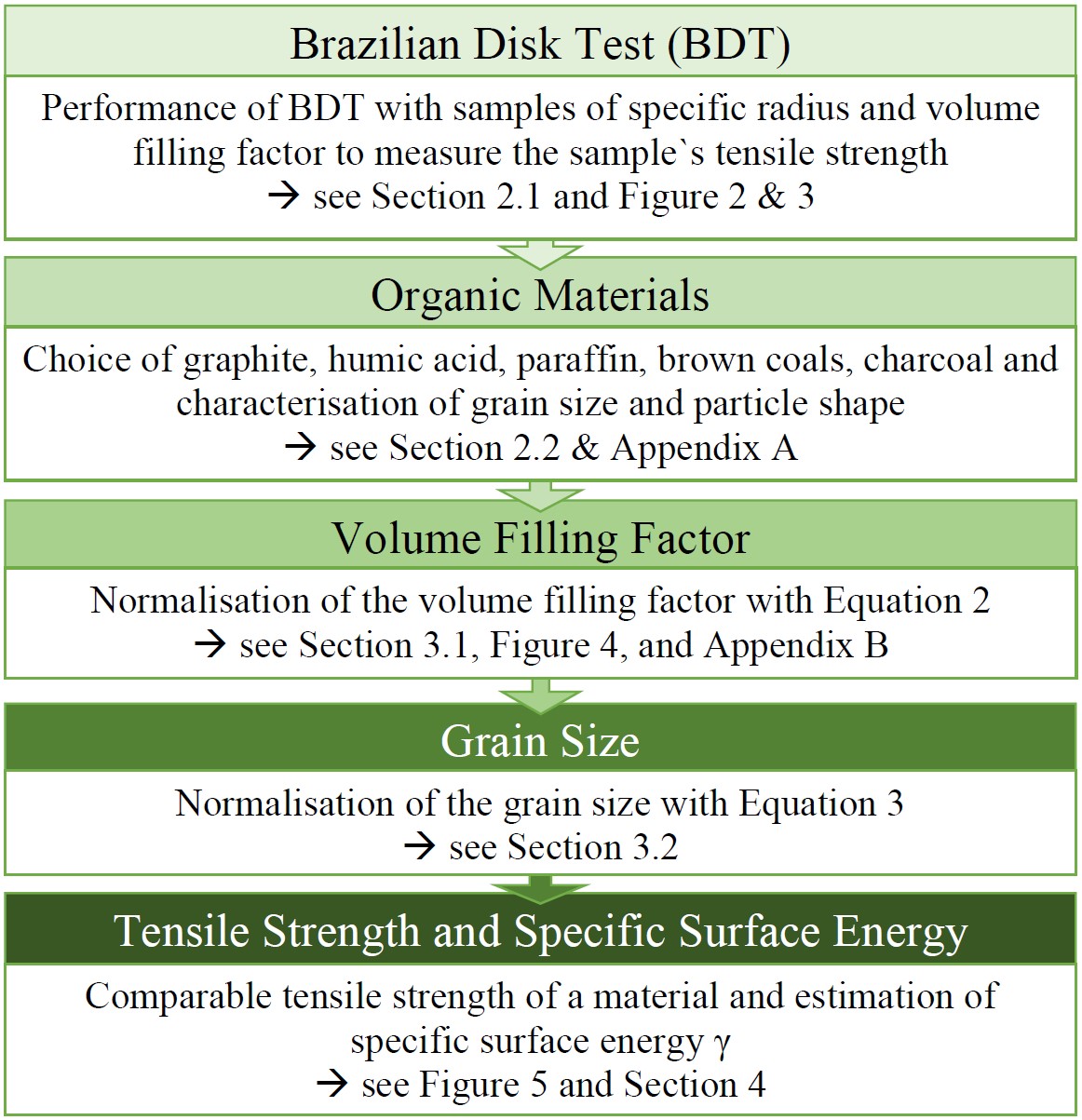}
    \caption{Overview of the procedure to estimate comparable tensile strength values for the selected organic materials.}
    \label{fig:procedure}
\end{figure}

Figure \ref{fig:procedure} summarises the method and logic of our work and references the respective Sections in this paper.

\subsection{Experimental Procedure}\label{Experimental_Procedure}
\begin{figure}
    \centering
    \includegraphics[width=\columnwidth]{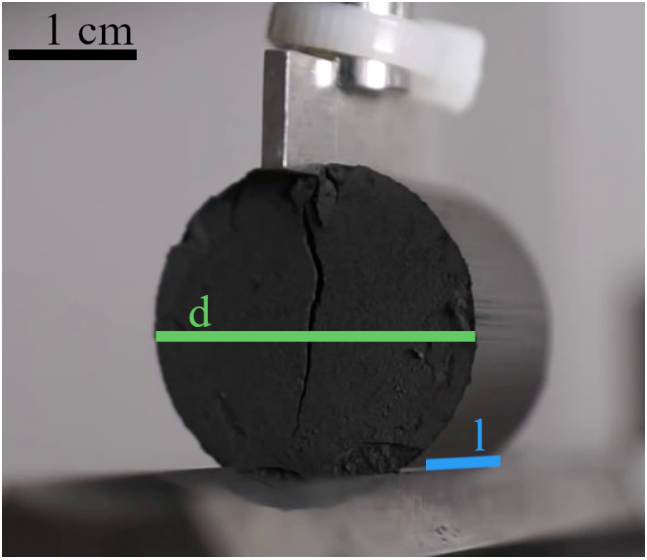}
    \caption{Geometry of the used Brazilian Disc Test setup. A loading platen is lowered onto a cylindrical sample by a stepper motor and exerts a force and, thus, a pressure onto the cylinder. The sample stays intact until the tensile strength of the material is reached. During breakup, a crack forms and the sample ideally is separated into two equally-sized pieces. The scale is not visible in the image and is located beneath the lower metal bar below the graphite sample.}
    \label{fig:measurment}
\end{figure}

\begin{figure}
    \centering
    \includegraphics[width=\columnwidth]{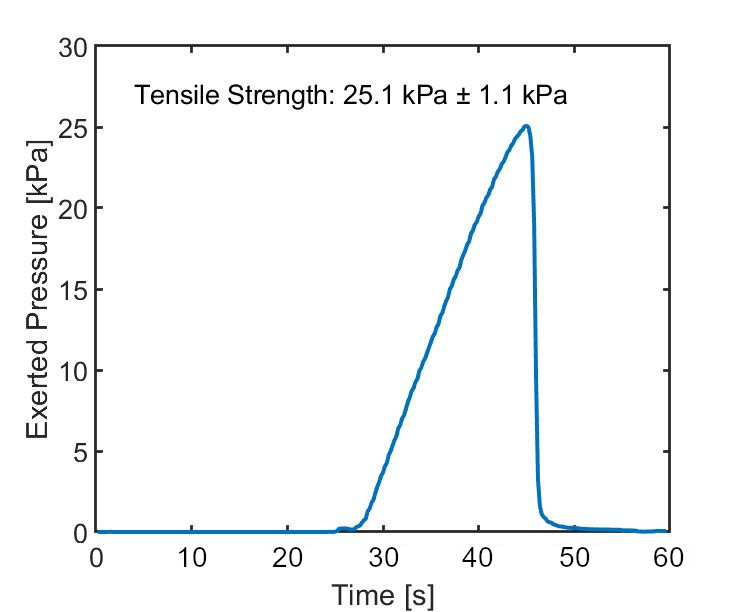}
    \caption{Example for the temporal evolution of the exerted pressure onto a cylindrical paraffin sample. The breakup of the cylinder was observed after $45 \,\mathrm{s}$ and the exerted pressure was $25.0 \,\mathrm{kPa} \pm 1.1\,\mathrm{kPa}$.}

    \label{fig:paraffin-example}
\end{figure}

In the following, we present the experimental setup used to study the tensile strength of micro-granular organic materials. As an indirect method, the Brazilian Disc Test (BDT) was used that is feasible to measure the tensile strength of solid samples and compressed granular matter. Since 1943, the BDT has been used with different solid materials and several publications deal with its applications \citep[see ][]{Li2013b}. The material must be shaped into a cylindrical form. In our case, the granular materials are filled into a cylindrical mould and are compressed by a piston \citep[see][for details]{Gundlachetal2018}. Then, a loading platen is used to exert a load onto the cylinder until it breaks. An example of a cylindrical sample under load is shown in Figure \ref{fig:measurment}. With the maximal exerted force of the material $F$ measured during breakup, the thickness of the sample $l$ (which varied in our experiments between $8$ and $24 \,\mathrm{mm}$) and the diameter of the sample (in our experiments $d = 25 \,\mathrm{mm}$), the tensile strength $\sigma$ can be derived through
\begin{equation}
    \sigma = \frac{2 \, F}{\pi \, d \, l} \, \mathrm{.}
    \label{eq:TS_grain_size_correction}
\end{equation}
The force is measured by a scale placed underneath the sample. With a stepper motor, the exerted force is continuously increased until the sample breaks. Figure \ref{fig:paraffin-example} shows a typical tensile-stress evolution of a paraffin-wax sample.

\subsection{Selected Materials}\label{Sample_Preparation}
For this study, several organic materials were selected, namely, paraffin wax, humic acid, graphite, three brown coals and one charcoal. Paraffin is a mixture of several alkane components, therefore hydrophobic, and was found in meteorites \citep{Nagy1961}. Humic acids are groups of complex organic acids and are main components of humus. The work of \citet{Sedletzky1935} revealed a chemical relation between humic acids, brown coals and graphite. Due to acidic and other groups, humic acid is partially hydrophilic. This applies also to the coals. However, the ratio of the hydrophobic part of the surface to the total surface, the hydrophobicity, can vary between less than 50\% to up to 95\% and depends on the exact material composition \citep{AdamsViola1981,Groszek1993,Xia2014}. For the materials used in our study, the hydrophobicity was not measured, but it should be kept in mind that hydrophilic parts can influence the tensile strength if water molecules stick to the particles and change their surface properties \citep{Kimura2015}. Because our measurements were performed at atmospheric conditions, it is not excluded that adsorbed water layers may influence the hydrophilic parts of the samples. For silica, this influence was highlighted by \citet{Kimura2015}. They found that the water layers at ambient pressure reduce the specific surface energy and therefore the tensile strength by a factor of up to 10 compared to measurements under vacuum conditions. Silica is highly hydrophilic (i.e., has a low hydrophobicity) so that we can assume, due to the higher hydrophobicity of humic acid and the coals, that the influence of adsorbed water in our study is less than this factor of 10. Because humic acids and coals are plant remains, they are not expected to be found in extra-terrestrial objects, but they can be potentially used as analogue materials. Furthermore, graphite was chosen, because it is an important analogue material, whose granular tensile strength was not studied before, and which has a fundamental affinity to the other organic materials, although it is an inorganic mineral. It is mostly hydrophobic with a hydrophobicity > 77\% \citep{Groszek1993} and therefore less effected by water as described above. Graphite was studied in two grain sizes, \textit{Graphite UF2} and \textit{Graphite AF}, respectively. Optical microscope images and macroscopic images of the samples as well as their size distributions can be found in Appendix \ref{sec:appendix_form}. In addition, Table \ref{tab:results} summarises the median radius, the median axis ratio and the median shape parameter of the materials used in our study. The median radius $\widetilde{r}$ is retrieved by two methods: 1) by the analysis of the cumulative normalised fraction by count (indicated in the following with the index $c$), and 2) by mass-weighting of the particles, assuming a spherical shape (indicated in the following with the index $m$). The axis ratio $C$ is defined as the ratio of semi-minor to semi-major axis, whereas the shape parameter $B$ is the area divided by the square of the perimeter normalised to the value of a circle. Details of the definition and the measurements of the axis ratio and shape parameter can also be found in Appendix \ref{sec:appendix_form}. From the distributions of the shape parameters, shown in Appendix \ref{sec:appendix_form}, the median and an uncertainty interval are retrieved and shown in \ref{tab:results}.
\begin{table*}
\centering
\caption{Summary of material properties and results for the tensile strength. The median grain radius $\widetilde{r}$ is shown with one standard deviation in parenthesis for the estimation by count ($\widetilde{r}_{c}$) and by mass ($\widetilde{r}_{m}$) (see Figures \ref{fig:GV_GraphiteAF} - \ref{fig:GV_BK_Vattenfall} in Appendix \ref{sec:appendix_form}). The median axis ratio $\widetilde{C}$ and the median shape parameter $\widetilde{B}$ indicate the form of the particles retrieved from their distributions, see Appendix \ref{sec:appendix_form} for details. The values for the tensile strength $\sigma_{c}$ and $\sigma_{m}$ are normalised to a grain radius of $r_0 =1 \,\mathrm{\mu m}$ and to a volume filling factor of $\Phi = 0.5$. The grain size normalisation was performed twice, following \citet{Gundlachetal2018}, shown in the upper line, and following \citet{Kimuraetal2020}, shown in the lower line for each material. In addition, the corresponding specific surface energies $\gamma_{c}$ and $\gamma_{m}$ are shown in the last column. The interval in brackets underneath equals the confidence interval of $1\sigma$ for the median, respectively.}
\label{tab:results}
\begin{tabular}{|c|c|c|c|c|c|c|c|c|}
\hline
Material  & $\widetilde{r}_{c} $ & $\widetilde{r}_{m} $  & $\widetilde{C} $ & $\widetilde{B} $ & ${\sigma}_{c} $ & ${\sigma}_{m} $ & ${\gamma}_{c} $ & ${\gamma}_{m} $ \\
 \hline
Unit & {$\mathrm{\mu m}$} & {$\mathrm{\mu m}$} & - & - &  {$\mathrm{Pa}$} & {$\mathrm{Pa}$} & $\mathrm{J/m^2} $ & $\mathrm{J/m^2} $\\
\hline
Graphite     &   0.89                & 8.3                  & 0.73                 & 0.46                 & $2.8 \cdot 10^4$                & $2.6 \cdot 10^5$             & $2.1 \cdot 10^{-1}$              & $1.9\cdot 10^{0}$ \\ 
(AF)         & \tiny{$(0.30-3.32) $} & \tiny{$(4.3-13.2) $} & \tiny{$(0.53-0.88)$} & \tiny{$(0.30-0.54)$} &  \tiny{$(0.9-10.4)\cdot 10^4$}  & \tiny{$(1.3-4.2)\cdot 10^5$} & \tiny{$(0.6-7.7) \cdot 10^{-1}$} & \tiny{ $(0.9-3.1) \cdot 10^{0}$}\\
             &                       &                      &                      &                      &  $3.0 \cdot 10^4$               & $9.1 \cdot 10^4$             & $2.2 \cdot 10^{-1}$              & $6.6 \cdot 10^{-1}$ \\
             &                       &                      &                      &                      &  \tiny{$(0.9-41.0)\cdot 10^4$}  & \tiny{$(7.3-10.8)\cdot 10^4$}& \tiny{$(0.6-3.0) \cdot 10^{-1}$} & \tiny{ $(5.3-7.9) \cdot 10^{-1}$}\\
\hline
Graphite     &   1.4                 & 3.4                  & 0.71                 & 0.49                 & $3.9 \cdot 10^5$                & $1.1 \cdot 10^6$             & $2.9 \cdot 10^0$                 & $8.2 \cdot 10^{0}$ \\ 
(UF2)        & \tiny{$(0.5-2.1) $}   & \tiny{$(2.4-5.9) $ } & \tiny{$(0.53-0.87)$ }& \tiny{$(0.37-0.56)$} & \tiny{ $(1.3-7.9)\cdot 10^5$ }  & \tiny{$(0.6-1.7)\cdot 10^6$} &\tiny{ $(0.9-5.8) \cdot 10^{0}$}  & \tiny{$(4.8-12.0) \cdot 10^{0}$} \\
             &                       &                      &                      &                      &  $3.3 \cdot 10^5$               & $5.6 \cdot 10^5$             & $2.4 \cdot 10^{0}$               & $4.1 \cdot 10^{0}$ \\
             &                       &                      &                      &                      &  \tiny{$(1.8-4.6)\cdot 10^5$}   & \tiny{$(4.1-6.9)\cdot 10^5$} & \tiny{$(1.3-3.4) \cdot 10^{0}$}  & \tiny{ $(3.1-5.1) \cdot 10^{0}$}\\
\hline
Humic Acid   &  14                   & 340                  & 0.64                 & 0.65                 & $4.9 \cdot 10^3$                & $1.2 \cdot 10^5$             & $3.6 \cdot 10^{-2}$              & $9.0 \cdot 10^{-1}$\\ 
(non-ground) & \tiny{$(6-53) $}      &\tiny{ $(260-390) $}  &\tiny{ $(0.46-0.82)$} & \tiny{ $(0.38-0.81)$}& \tiny{ $(0.7-19.2)\cdot 10^3$ } &\tiny{ $(0.4-2.1)\cdot 10^5$ }& \tiny{$(0.5-14.0) \cdot 10^{-2}$}&\tiny{ $(2.8-15.0) \cdot 10^{-1}$}  \\
             &                       &                      &                      &                      &  $1.3 \cdot 10^3$               & $6.6 \cdot 10^3$             & $9.8 \cdot 10^{-3}$              & $4.9 \cdot 10^{-2}$ \\
             &                       &                      &                      &                      &  \tiny{$(0.5-2.2)\cdot 10^3$}   & \tiny{$(2.3-11.0)\cdot 10^3$}& \tiny{$(3.4-16.1) \cdot 10^{-3}$}& \tiny{ $(1.7-8.0) \cdot 10^{-2}$}\\
\hline
Humic Acid   &   0.75                & 2.7                  & 0.73                 & 0.48                 & $2.3 \cdot 10^2$                & $8.3 \cdot 10^2$             & $1.7 \cdot 10^{-3}$              & $6.1 \cdot 10^{-3}$ \\ 
(ground)     & \tiny{$(0.27-2.69) $} & \tiny{$(1.2-5.4) $}  & \tiny{$(0.56-0.85)$} & \tiny{$(0.32-0.62)$} & \tiny{ $(0.2-8.5)\cdot 10^2$ }  & \tiny{$(0.9-18.2)\cdot 10^2$}& \tiny{$(0.1-6.2) \cdot 10^{-3}$} & \tiny{$(0.6-13.4) \cdot 10^{-3}$ }\\
             &                       &                      &                      &                      &  $2.7 \cdot 10^2$               & $5.0 \cdot 10^2$             & $2.0 \cdot 10^{-3}$              & $3.7 \cdot 10^{-3}$ \\
             &                       &                      &                      &                      &  \tiny{$(0.2-8.5)\cdot 10^2$}   & \tiny{$(1.4-8.6)\cdot 10^2$} & \tiny{$(0.4-3.5) \cdot 10^{-3}$} & \tiny{ $(1.0-6.3) \cdot 10^{-3}$}\\
\hline 
Paraffin     & 260                   & 500                  & 0.92                 & 0.76                 & $5.3 \cdot 10^5$                & $1.0 \cdot 10^6$             & $3.9 \cdot 10^0$                 & $7.4 \cdot 10^{0}$\\ 
             & \tiny{$(70-330) $}    & \tiny{$(380-570) $}  & \tiny{$(0.68-0.98)$ }& \tiny{$(0.62-0.86)$} &  \tiny{$(1.3-6.6)\cdot 10^5$ }  & \tiny{$(0.7-1.2)\cdot 10^6$} & \tiny{$(1.0-4.8) \cdot 10^{0}$}  & \tiny{$(5.4-8.7) \cdot 10^{0}$} \\
             &                       &                      &                      &                      &  $3.3 \cdot 10^4$               & $4.5 \cdot 10^4$             & $2.4 \cdot 10^{-1}$              & $3.3 \cdot 10^{-1}$ \\
             &                       &                      &                      &                      &  \tiny{$(2.9-3.6)\cdot 10^4$}   & \tiny{$(3.9-5.0)\cdot 10^4$} & \tiny{$(2.1-2.7) \cdot 10^{-1}$} & \tiny{ $(2.9-3.7) \cdot 10^{-1}$}\\
\hline
Brown Coal   &   0.87                & 12                   & 0.79                 & 0.55                 & $2.0 \cdot 10^2$                & $2.8 \cdot 10^3$             & $1.5 \cdot 10^{-3}$              & $2.0 \cdot 10^{-2}$\\ 
Romonta      & \tiny{$(0.16-2.09) $} & \tiny{$(6-13) $ }    & \tiny{$(0.60-0.90)$} & \tiny{$(0.37-0.63)$} & \tiny{$(0.1-5.5)\cdot 10^2$}    & \tiny{$(0.4-5.4)\cdot 10^3$} & \tiny{$(0.1-4.0) \cdot 10^{-3}$}& \tiny{$(0.3-4.0) \cdot 10^{-2}$}  \\
             &                       &                      &                      &                      &  $2.2 \cdot 10^2$               & $8.1 \cdot 10^2$             & $1.6 \cdot 10^{-3}$              & $5.9 \cdot 10^{-3}$ \\
             &                       &                      &                      &                      &  \tiny{$(0.2-4.5)\cdot 10^2$}   & \tiny{$(1.9-15.9)\cdot 10^2$}& \tiny{$(0.2-3.3) \cdot 10^{-3}$} & \tiny{ $(1.4-11.6) \cdot 10^{-3}$}\\
\hline
Brown Coal   &   1.1                 & 8.4                  & 0.74                 & 0.53                 & $7.1 \cdot 10^1$                & $5.4 \cdot 10^2$             & $5.2 \cdot 10^{-4}$              & $4.0 \cdot 10^{-3}$\\ 
RWE          & \tiny{$(0.4-3.2) $}   & \tiny{$(4.2-13.0) $} & \tiny{$(0.55-0.86)$ }& \tiny{$(0.34-0.64)$} &  \tiny{$(.1-22.1)\cdot 10^1$}   & \tiny{$(1.2-11.0)\cdot 10^2$}& \tiny{$(0.8-16.2) \cdot 10^{-4}$}&\tiny{ $(0.1-8.1) \cdot 10^{-3}$} \\
             &                       &                      &                      &                      &  $6.8 \cdot 10^1$               & $1.9 \cdot 10^2$             & $5.0 \cdot 10^{-4}$              & $1.4 \cdot 10^{-3}$ \\
             &                       &                      &                      &                      &  \tiny{$(0.1-13.0)\cdot 10^1$}  & \tiny{$(0.6-3.5)\cdot 10^2$} & \tiny{$(0.8-9.5) \cdot 10^{-4}$} & \tiny{ $(0.4-2.6) \cdot 10^{-3}$}\\
\hline
Brown Coal   &   1.0                 & 9.4                  & 0.80                 & 0.56                 & $8.2 \cdot 10^1$                & $7.4 \cdot 10^2$             & $6.0 \cdot 10^{-4}$              & $5.4 \cdot 10^{-3}$\\ 
Vattenfall   & \tiny{$(0.4-2.6) $}   & \tiny{$(5.0-11.4) $ }& \tiny{$(0.65-0.90)$} & \tiny{$(0.44-0.66)$} &  \tiny{$(1.5-22.2)\cdot 10^1$}  & \tiny{$(1.8-13.3)\cdot 10^2$}& \tiny{$(1.1-16.3) \cdot 10^{-4}$}& \tiny{$(1.3-9.8) \cdot 10^{-3}$}\\
             &                       &                      &                      &                      &  $8.1 \cdot 10^1$               & $2.4 \cdot 10^2$             & $5.9 \cdot 10^{-4}$              & $1.8 \cdot 10^{-3}$ \\
             &                       &                      &                      &                      &  \tiny{$(1.4-15.0)\cdot 10^1$}  & \tiny{$(1.0-4.3)\cdot 10^2$} & \tiny{$(1.0-11.0) \cdot 10^{-4}$}& \tiny{ $(0.7-3.2) \cdot 10^{-3}$}\\
\hline
Charcoal     &   1.2                 & 5.1                  & 0.78                 & 0.52                 & $8.2 \cdot 10^2$                & $3.5 \cdot 10^3$             & $6.2 \cdot 10^{-3}$              & $2.5 \cdot 10^{-2}$\\ 
             & \tiny{$(0.5-2.7) $}   & \tiny{$(2.7-6.9) $}  & \tiny{$(0.59-0.88)$} & \tiny{$(0.40-0.61)$} &  \tiny{$(2.8-18.4)\cdot 10^2$}  & \tiny{$(1.4-5.1)\cdot 10^3$} & \tiny{$(2.1-13.5) \cdot 10^{-3}$}& \tiny{$(1.1-3.8) \cdot 10^{-2}$} \\
             &                       &                      &                      &                      &  $7.5 \cdot 10^2$               & $1.5 \cdot 10^3$             & $5.5 \cdot 10^{-3}$              & $1.1 \cdot 10^{-2}$ \\
             &                       &                      &                      &                      &  \tiny{$(2.4-11.2)\cdot 10^2$}  & \tiny{$(0.9-2.1)\cdot 10^3$} & \tiny{$(1.7-8.2) \cdot 10^{-3}$} & \tiny{ $(0.7-1.5) \cdot 10^{-2}$}\\
\hline
\end{tabular}
\end{table*}

\section{Results}\label{Sect_3}
In total, 75 Brazilian Disc Tests were performed in this study and the results of these measurements are presented hereafter. In order to provide comparable values for the tensile strengths of the different materials, we  performed two normalisations to the measured data, which are explained in the following: a) volume-filling-factor normalisation and b) grain-size normalisation.

\vspace{0.2cm}
\par\noindent
\textbf{a) Volume-Filling-Factor Normalisation}
\par
\begin{figure*}
    \centering
    \begin{minipage}[b]{0.49\textwidth}
    \includegraphics[width=\columnwidth]{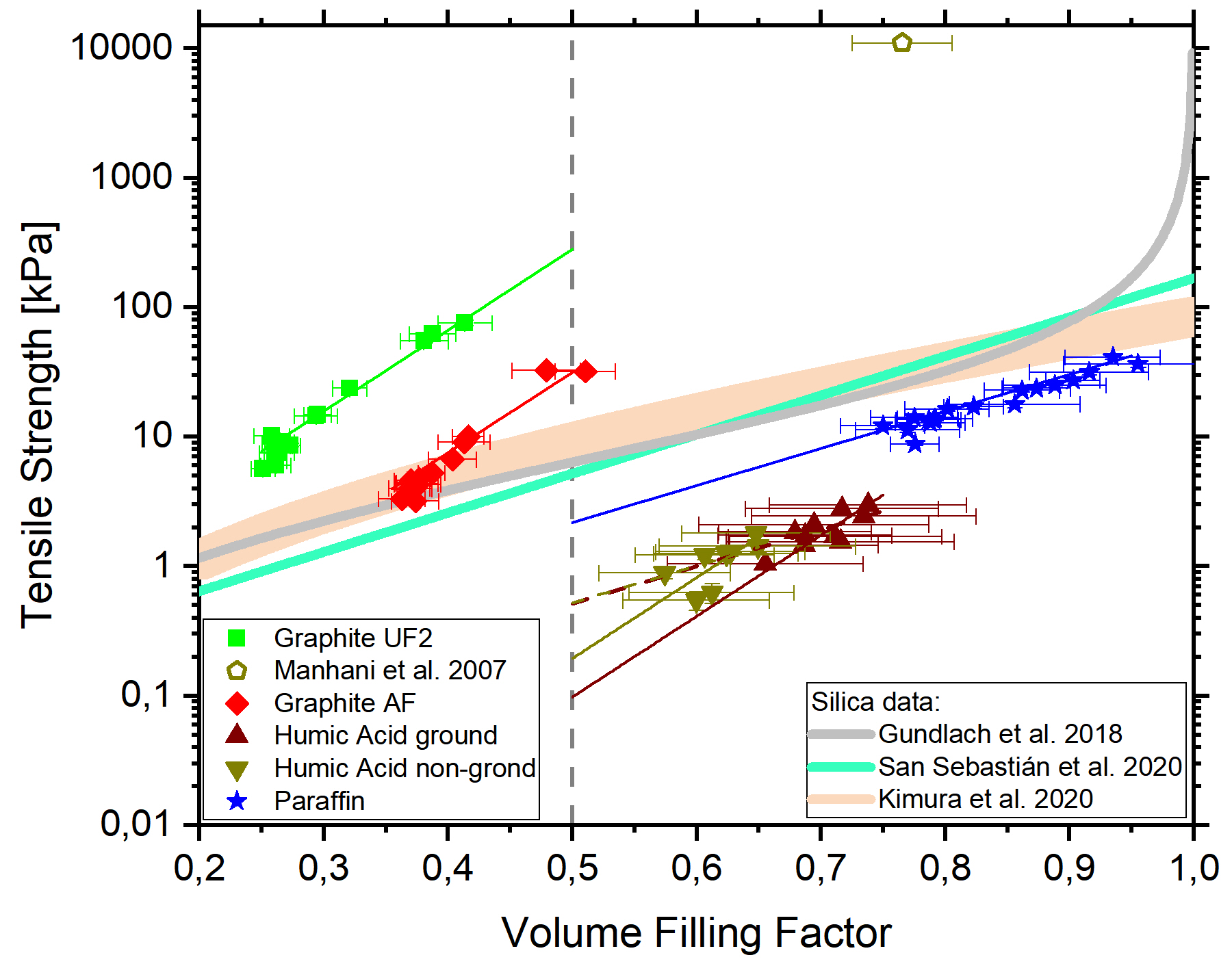}
    \end{minipage}
    \begin{minipage}[b]{.49\textwidth}
    \includegraphics[width=\columnwidth]{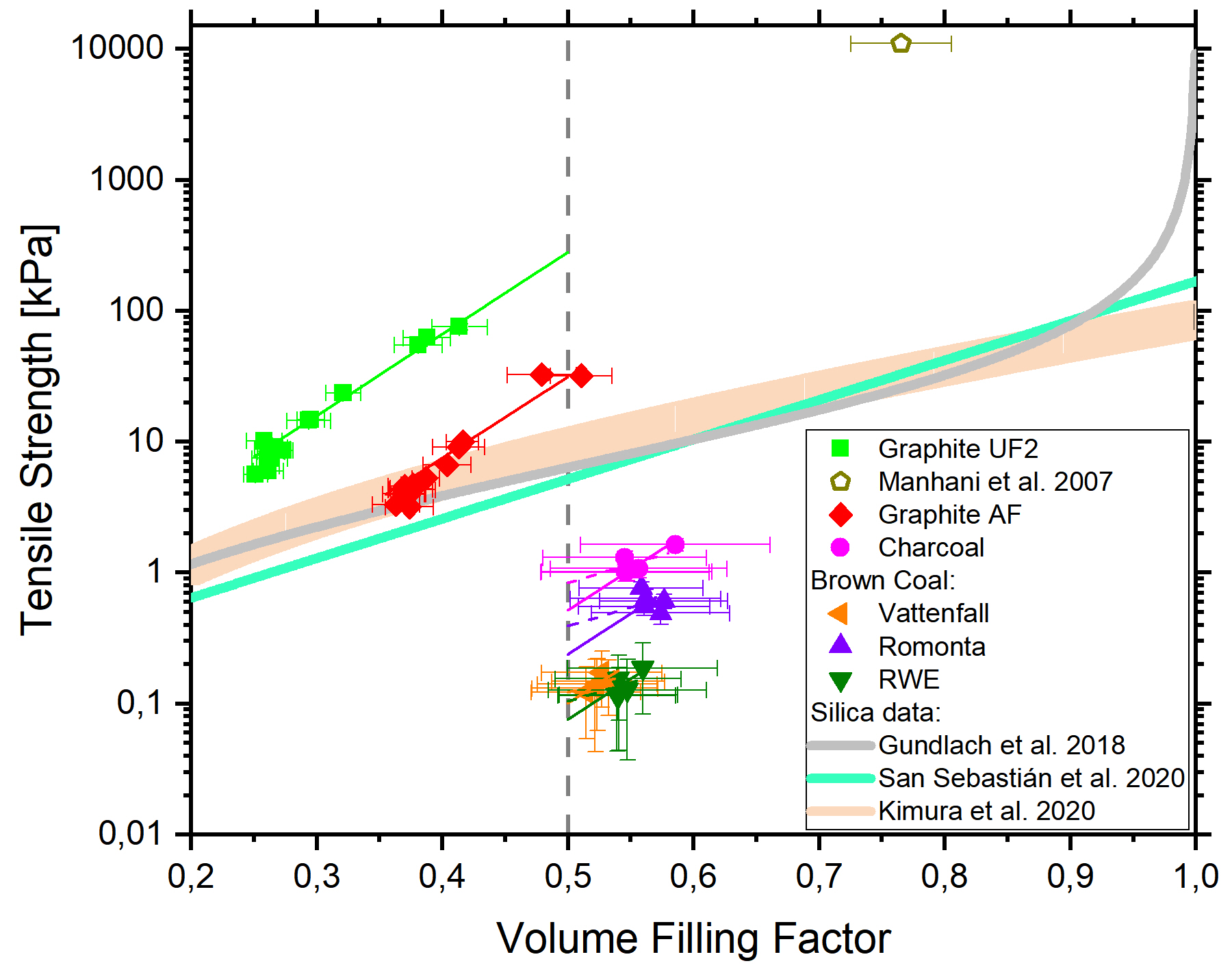}
    \end{minipage}
    \caption{Results of the Brazilian Disc Tests as a function of the volume filling factor of the samples for graphite, humic acid and paraffin (left) and for the coals (right, with the graphite data for comparison) with the fitting functions according to Equation \ref{eq:VFF_Fit} for the volume filling factor normalisation. For humic acid and the coals, the slope $a$ was set to values of 6.3 and 2.9, respectively, as indicated with the solid and dashed lines. For comparison, three functions for silica are plotted, namely from \citet{Gundlachetal2018}, \citet{Kimuraetal2020} and \citet{SanSebastian2020}, using the fits from \citet{SanSebastian2020}. The open pentagon indicates a data point for graphite from \citet{manhani200} with a median grain size of $838\, \mathrm{\mu m}$.}
    \label{fig:result_fit_1}
\end{figure*}

\noindent
It is obvious that the volume filling factor $\Phi$ significantly influences the tensile strength (see Figure \ref{fig:result_fit_1}). An increase of the volume filling factor increases the tensile strength, because the number of neighbouring particles in contact is enhanced. Graphite shows the strongest effect, but all samples exhibit a significant increase of the tensile strength with increasing volume filling factor. In the literature, several relations for this dependency are proposed \citep[see, e.g.][]{Meisner2012,Gundlachetal2018,Kimuraetal2020}. We tested the different functions and found that 
\begin{equation}
    \label{eq:VFF_Fit}
    \sigma = 10^{a(\Phi - b)} 
\end{equation}
fits our data best. The parameters $a$ and $b$ are obtained by least-squares-fitting Equation \ref{eq:VFF_Fit} to the data. For both humic acid data sets and all coals, two fits were performed with fixed $a=6.3$ (corresponding to the highest slope value for graphite; solid lines in Figure \ref{fig:result_fit_1}) and $a=2.9$ (corresponding to the lowest slope value for paraffin; dashed lines in Figure \ref{fig:result_fit_1}), because these data sets did not show enough variation in volume filling factor for a robust two-parameter fit. A summary of all obtained fitting parameters and the goodness of the fits can be found in Appendix \ref{appendix:vff_fit_parameter}. By using the best-fitting function, we then derived the tensile strength for a volume filling factor of $\Phi = 0.5$ for all the samples.  However, the parameter $b$ was adapted accordingly. Fortunately, the volume filling factors of the coal samples after compression were relatively close to $\Phi = 0.5$, so that the extrapolation was performed for only small changes in volume filling factor of $\Delta \Phi < 0.1$.
\par
For comparison, three functions of the volume-filling-factor dependency for silica from \citet{Gundlachetal2018}, \citet{Kimuraetal2020} and \citet{SanSebastian2020} are also plotted in Figure \ref{fig:result_fit_1}. They show similar slopes as paraffin, which has the smallest slope among the used materials in this study. From \citet{manhani200} an additional data point for graphite was retrieved (open pentagon in Figure \ref{fig:result_fit_1}). They used a highly compacted graphite sample with a median grain size of $838 \,\mathrm{\mu m}$ whose volume filling factor was around 0.77 and measured a tensile strength of $11.7 \,\mathrm{MPa}$. The data point of \citet{manhani200} indicates that the high slope for our graphite samples is real, although, due to the extremely different grain sizes, an extrapolation from our data to those of \citet{manhani200} is a difficult task (see below). The exact reason for the much higher slope of the graphite data compared to the other materials, including silica, remains unclear. One could speculate that for graphite, the volume filling factor depends on the coordination number in a different way as for the other materials. It should be mentioned that \citet{tsubaki1984} found comparably high slopes for carbonaceous materials, namely lactose and several coals.

\vspace{0.2cm}
\par\noindent
\textbf{b) Grain-Size Normalisation}
\par\noindent
\citet{Gundlachetal2018} showed that the tensile strength of granular matter consisting of micrometric spherical SiO$_2$ particles is inversely proportional to the grain radius. However, also other scaling relations have been proposed, e.g. by \citet{Kimuraetal2020}. Their work indicates an inverse proportionality to the square root of the particle radius under the condition that the Weibull modulus is six, valid for carbonaceous matter. Hence, the next step required for a proper comparison of the results is a normalisation of the grain size. Therefore, the equations
\begin{equation}
    \sigma_0 =\sigma_r \, \frac{\widetilde{r}}{r_0} \, ,
    \label{eq:TSN1}
\end{equation}
after \citet{Gundlachetal2018}, and 
\begin{equation}
    \sigma_0 =\sigma_r \, \sqrt{\frac{\widetilde{r}}{r_0}} \, ,
    \label{eq:TSN2}
\end{equation}
after \citet{Kimuraetal2020}, are used in the following. Here, $r_0$ is the reference grain radius of $1 \, \mathrm{\mu m}$, $\widetilde{r}$ is the median radius of the size distribution, $\sigma_r$ is the measured tensile strength corrected to $\phi = 0.5$, and $\sigma_0$ is the tensile strength normalised to a grain radius of $1 \, \mathrm{\mu m}$ and a filling factor of $\phi = 0.5$, respectively. It should be mentioned that the particle shape might also have an influence on the tensile strength, which, however, we have not taken into account in this work.

\vspace{0.2cm}
\par\noindent
\textbf{Comparison of the Tensile Strength Values}
\par
\begin{figure*}
    \centering
    \includegraphics[width=\textwidth]{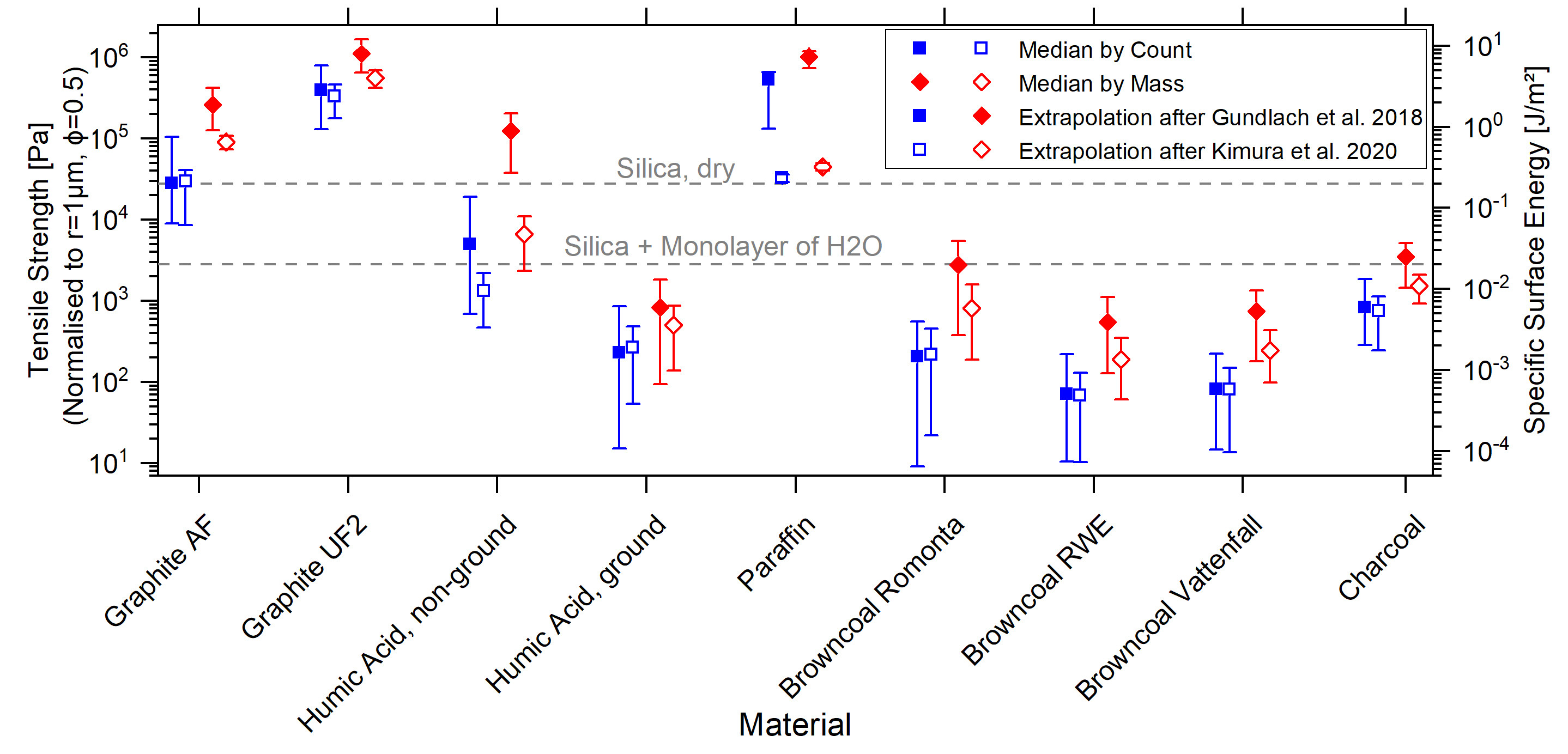}
    \caption{Normalised tensile-strength values $\sigma(\Phi = 0.5, r_0 =1 \,\mathrm{\mu m})$ for all investigated samples. Squares: normalised values calculated using the median radius estimated by the size-frequency distribution. Diamonds: normalised values calculated using the median radius estimated by the mass-frequency distribution. Filled symbols: size normalisation after Eq. \ref{eq:TSN1} \citep{Gundlachetal2018}. Open symbols: size normalisation after Eq. \ref{eq:TSN2}  \citep{Kimuraetal2020}. As a reference, values for silica are shown by the horizontal dashed lines. The upper line indicates the tensile strength of dry silica and the lower one corresponds to the tensile strength of silica coated with a monolayer of water \citep{Gundlachetal2018,Kimura2015,Steinpilz2019}. Furthermore, the derived specific surface energies are shown by the right y axes.}
    \label{fig:finalresult}
\end{figure*}

\noindent
The tensile strength values normalised to a standard volume filling factor of $\Phi = 0.5$ and a standard grain radius of $1 \, \mathrm{\mu m}$ of the analysed sample materials are summarised in Figure \ref{fig:finalresult} and Table \ref{tab:results}. A comparison to our previous work \citep{Gundlachetal2018} is provided by the dashed grey lines, visualising the tensile strength of dry and wet silica samples. 

For grain sizes $ \gg 1 \,\mathrm{\mu m}$ the two size corrections after Eqs. \ref{eq:TSN1} and \ref{eq:TSN2} should lead to notable differences in the normalised tensile strengths. However, in our samples this is, for the median grain radius determined through the size-frequency distribution, only the case for paraffin. For all other samples, these differences are within the error margins. The differences between the two normalisation methods become more pronounced for the cases in which we determined the median grain radius through the mass-frequency distribution. Here, graphite AF, non-ground humic acid and paraffin show differences outside the error bars. However, the general material dependency of the tensile strength remains the same for the two normalisation cases. 

Both graphite samples and paraffin possess a much higher tensile strength as silica. All other tested materials are either comparable to silica (humic acid, non-ground), or have lower tensile strength values (humic acid, ground; charcoal and all brown coals). 

The difference in tensile strength of the ground and the non-ground humic acid is surprising. As we normalise to a standard grain size, the grinding state should actually not influence the normalised tensile strength. However, we can speculate that the measured difference is caused either by the change of the size-frequency distribution (see Figures \ref{fig:GV-Huminsaure-ungem} and \ref{fig:GV-Huminsaure-gem}) or of the shape of the particles. A closer look to the size- and mass-frequency distributions shows some differences. First, the size-frequency distributions of the non-ground and ground material are very similar in shape, whereas the mass-frequency distributions differ strongly. The mass-frequency distribution of the non-ground humic acid is dominated by a few large particles so that the curve gets very steep and small particles play only a negligible role. This leads to a possible higher uncertainty in the normalisation of the grain-size and, therefore, the tensile-strength value estimated with the mass-weighted median radius is questionable and possibly over-estimated. Second, the shape-parameter distributions (see \ref{fig:Form_H_P}) reveal a clear difference between the ground and non-ground case. The ground humic acid seems to have more compact particles, because the axis ratio is higher than for non-ground humic acid, but the shape parameter $B$ is decreased compared to the non-ground material. This means, that the ground particles have more edges, whereas the non-ground grains seem to be rounder. Regarding the difference in tensile strength, this leads to the speculation that edges on particles lower the tensile strength of the bulk material.

Furthermore, also the two graphites show a discrepancy after normalisation. Their shape parameters are broadly in agreement, but regarding the size-frequency distribution, graphite UF2 seems to be less polydisperse, which could lead to a higher tensile strength. It should be noted here that the manufacturer provides different size-ranges compared to our measurements, in which graphite UF2 is smaller than AF (UF2: $4.5^{+3.8}_{-2.4}\,\mathrm{\mu m} $, AF: $9.5^{+12.0}_{-6.1}\,\mathrm{\mu m} $; Graphit Kropfm\"uhl GmbH). Compared with literature values, the specific surface energy of graphite AF matches pretty well the results of earlier studies  \citep[the values range between $0.063 \pm 0.007 \,\mathrm{J/m^2}$ \citep{Ferguson2016} and $0.260 \pm 0.029 \,\mathrm{J/m^2}$ \citep{Girifalco1956}, see also][]{Han2019}. However, as mentioned before, \citet{manhani200} measured the tensile strength of a compact graphite sample whose volume filling factor is about 0.77. Their result of a tensile strength of $11.7 \,\mathrm{MPa}$ falls in the regime we would expect, but due to the different grain size with a median of $838 \,\mathrm{\mu m}$, we cannot extrapolate our data to it. 
Tensile strength values depending on the porosity for lactose with a particle diameter of $46.9 \,\mathrm{\mu m}$  were presented by \citet{tsubaki1984} and vary between $\sim 10 \,\mathrm{Pa} \, (\Phi \approx 0.3)$ and $\sim 1000 \,\mathrm{Pa} \,(\Phi \approx 0.58)$. These values are comparable to the results of the coals and the slope matches broadly the graphite slope. \citet{tsubaki1984} also performed experiments with several coals resulting in  similar slopes, but, due to their data presentation, exact values cannot be compared with our results.

Similarly, one can compare the coals in a more detailed way. A significant difference in the shape parameters among the brown coals and the charcoal is not present, but the charcoal possesses a significantly higher tensile strength. Thus, two other factors can possibly be responsible for the difference in tensile strength. Either the width of the size distribution of charcoal, which is narrower than those of the brown coals, enhances the tensile strength, or an intrinsic difference in the material could be responsible for the higher tensile strength.

In summary, there are indications that a strong dependency of the tensile strength on particle shape and size-distribution width exists, which requires further investigations.

Since the tensile strength is correlated with the specific surface energy of the material \citep[see Equation 2 in][]{Gundlachetal2018}, the corresponding values of the specific surface energies of all studied samples are shown on the right-hand y-axes in Figure \ref{fig:finalresult} as well as in the Table \ref{tab:results}. The uncertainty of the tensile strength is calculated by taking into account the uncertainties of the volume filling factor fit and of the mean grain size determination. 

\section{Discussion and applications to comets}\label{Sect_4}
As can be seen in Figure \ref{fig:finalresult}, organic materials are not generally stickier than other materials, like silica, or water ice. The spread of the normalised tensile-strength values ranges from very low $(7.1^{+15.0}_{-7.0} ) \times 10^{1} \, \mathrm{Pa}$ (browncoal RWE) to very high values $(1.1^{+0.6}_{-0.5} ) \times 10^6 \, \mathrm{Pa}$ (graphite UF2). This shows that the tensile strength strongly depends on the material, the size-frequency distribution and the shape of the grains. Thus, the class of organic materials does not \textit{per se} possess only extraordinarily high tensile-strength values.
\par
In cometary science, organic materials are often considered as a \grqq glue\grqq\ between the particles, the agglomerates, and the pebbles, which leads to a significant increase of the tensile strength of the cometary surface. Since dust emission from a cometary surface can only work if the gas pressure exceeds the tensile strength of the material \citep[see, e.g.,][]{Fulleetal2019,Fulleetal2020a,Fulleetal2020b,Gundlachetal2020}, the presence of organic matter with high tensile-strength values would lead to the suppression of cometary dust ejection. However, this work shows that cometary activity can be maintained, or even eased also when organic materials are present. Only a high abundance of graphite-like organics would quench cometary activity. This finding is in agreement with laboratory measurements performed by \citet{Zhang2020}, who found a higher sticking velocity for graphite particles compared to silica particles in their collision experiments. The influence of paraffin is not discussed here, because it is a synthetic product.
\par 
The variable tensile strength of organic materials also has an impact on planet formation theories. For example, \citet{Homma_2019} focused on the shear and Young`s modulus of the material to estimate their stickiness and therefore they assumed equal specific surface energy for the silica and the organic materials. Because of the relation between the tensile strength and the specific surface energy, our work shows that this assumption is only valid for a specific organic material, namely for humic acid (non-ground; mean radius estimated through the size-frequency distribution). Hence, it would be interesting to include the other organics in this theory to understand how the change of the specific surface energy influences the shear and Young's modulus.
\par
Our data indicate that more complex organic materials (humic acid, coal) possess a lower tensile strength than simpler organic materials (graphite, humic acid). This seems to be in contrast to the findings of \citet{kimura2020b} that increasing carbonisation decreases the impact velocity for fragmentation. The latter should be proportional to the square root of the surface energy, which, in turn, is proportional to the tensile strength  \citep{SanSebastian2020}. However, as shown by \citet{Gundlach2015b} with the comparison of silica and water ice, the surface energy is not the only value that determines the sticking or fragmentation strength.
\par
Furthermore, the results of this paper emphasise the need for a cometary sample-return mission, since the exact knowledge of the chemical composition and the morphology of the cometary grains are mandatory to understand the physics of the cometary dust-ejection mechanisms and for the preparation of high-quality cometary analogue materials. It should be noted here that samples from the interior of a comet are required, due to the possible processing of the surface material \citep{BockeleeMorvan2019}. 
\par
\citet{Bardynetal2017} argued, based on the O/C ratio of $\sim0.2$ of the cometary dust measured by the COSIMA instrument, that the organic compound can be best compared to insoluble organic matter found in carbonaceous chondrites. Insoluble organic matter has a rather complex structure \citep{REMUSATetal2007} and this is why we foresee to use such materials for our future CoPhyLab experiments.

\section{Conclusion}\label{Sect_5}
We measured the tensile strength of different organic materials with the Brazilian Disc Test. The main findings of this work are that
\begin{enumerate}
\item organic materials are not generally stickier than other materials, but may, on the opposite, even be less sticky than silicates,
\item graphite and paraffin possess very high tensile-strength values of $(2.8^{+7.6}_{-1.9} ) \times 10^4 \, \mathrm{Pa}$ - $(1.1^{+0.6}_{-0.5} ) \times 10^6 \, \mathrm{Pa}$ (graphite) and $(3.3^{+0.3}_{-0.4} ) \times 10^4 \, \mathrm{Pa}$ - $(1.0^{+0.2}_{-0.3} ) \times 10^6 \, \mathrm{Pa}$ (paraffin),
\item all tested coals have very low tensile-strength values, ranging from $(7.1^{+15.0}_{-7.0} ) \times 10^{1} \, \mathrm{Pa}$ to $(3.5^{+1.6}_{-2.1} ) \,\times 10^{3} \mathrm{Pa}$, and 
\item only humic acid shows tensile-strength values comparable to silica, i.e., $(2.3^{+6.2}_{-2.1} ) \times 10^{2} \, \mathrm{Pa}$ to $(1.2^{+0.9}_{-0.8} ) \times 10^{5} \, \mathrm{Pa}$ .
\end{enumerate}
Our data indicate a trend that more complex organic materials possess lower tensile strengths than simpler organics. These results underline even more that we need a precise determination of the cometary material composition to understand the influence of the organic compounds on the dust activity of comets. The Comet Interceptor mission, to be launched in 2028, will hopefully help to estimate the material composition of a cometary surface.

\section*{Acknowledgements}
This work was carried out in the framework of the CoPhyLab project funded by the D-A-CH programme (GU 1620/3-1 and BL 298/26-1 / SNF 200021E 177964 / FWF I 3730-N36). DB and JB thank the Deutsches Zentrum f\"ur Luft- und Raumfahrt for support under grant 50WM1846.

\section*{Data Availability}
The data underlying this article will be shared on reasonable request to the corresponding author.

\bibliographystyle{mnras}
\bibliography{references}

\appendix

\section{Size Distribution and Shape of Particles}
\label{sec:appendix_form}

\begin{figure*}
    \centering
    \begin{minipage}[b]{0.45\textwidth}
    \includegraphics[width=\columnwidth]{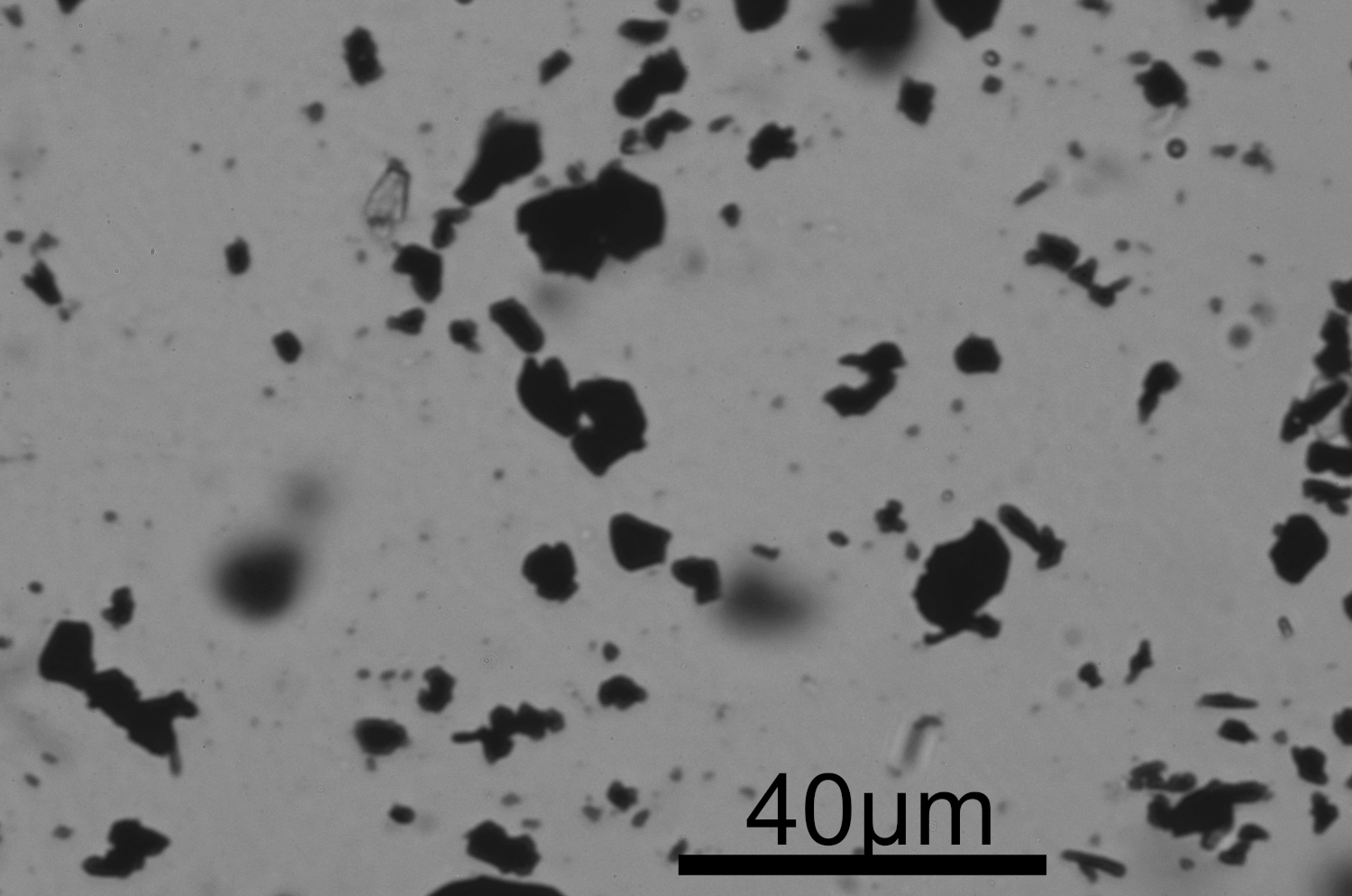}
    \end{minipage}
    \begin{minipage}[b]{.45\textwidth}
    \includegraphics[width=\columnwidth]{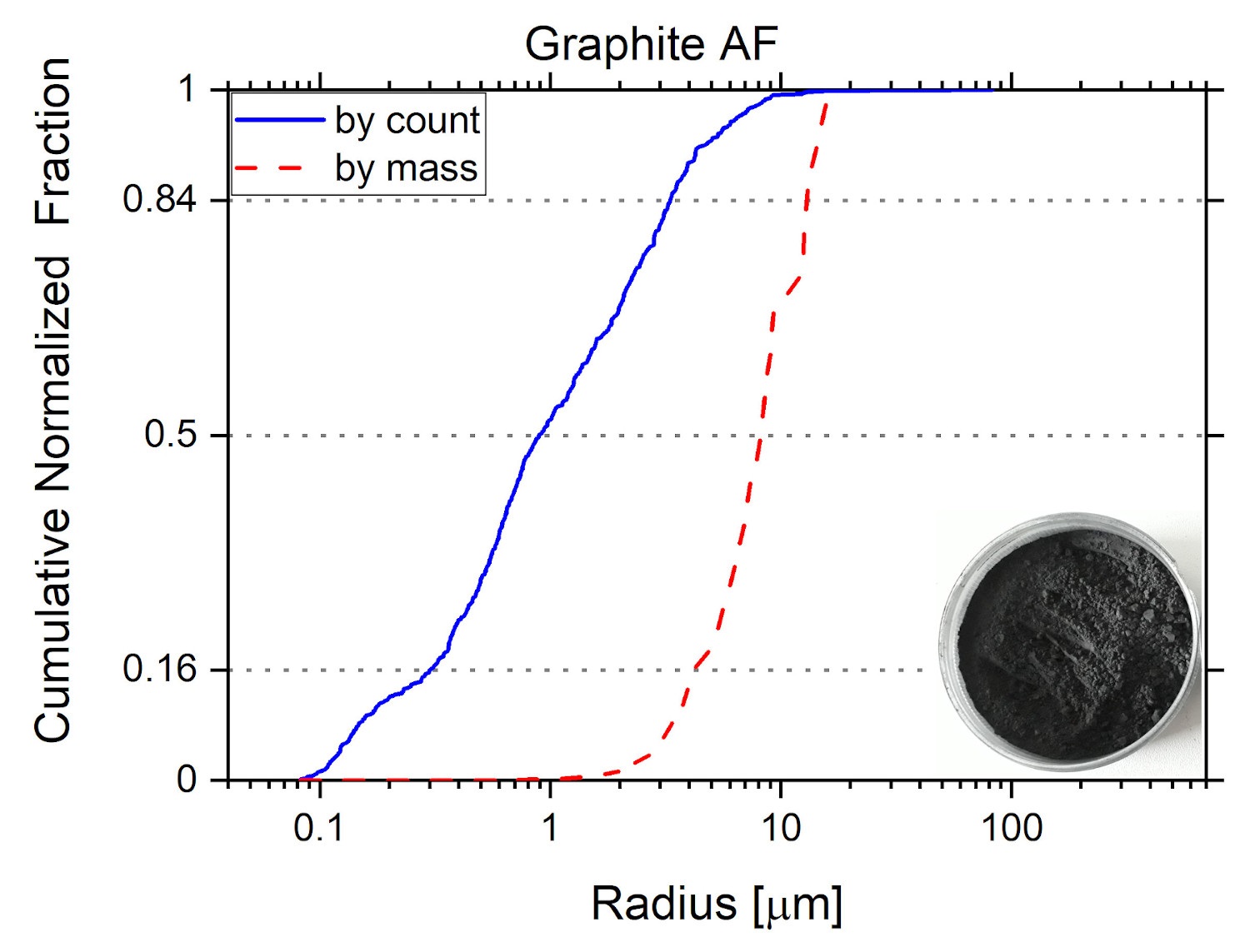}
    \end{minipage}
    \caption{Left: Microscopic caption of Graphite AF. Right: Size distributions of Graphite AF calculated by count and by mass. The median and the $1 \sigma$ interval is indicated by the dotted lines. A macroscopic picture of the sample material is also shown.}
    \label{fig:GV_GraphiteAF}
\end{figure*}

\begin{figure*}
    \centering
    \begin{minipage}[b]{0.45\textwidth}
    \includegraphics[width=\columnwidth]{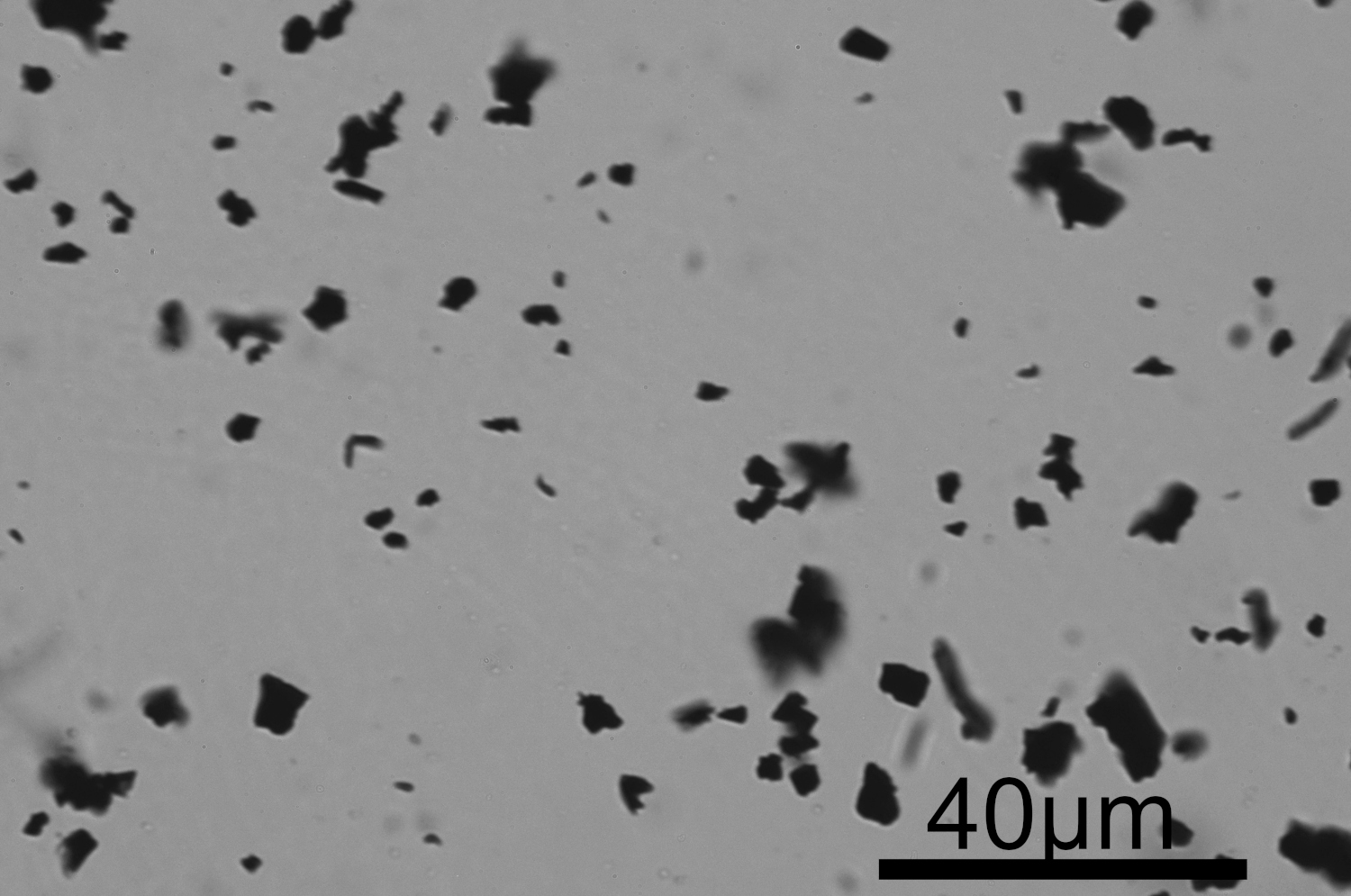}
    \end{minipage}
    \begin{minipage}[b]{.45\textwidth}
    \includegraphics[width=\columnwidth]{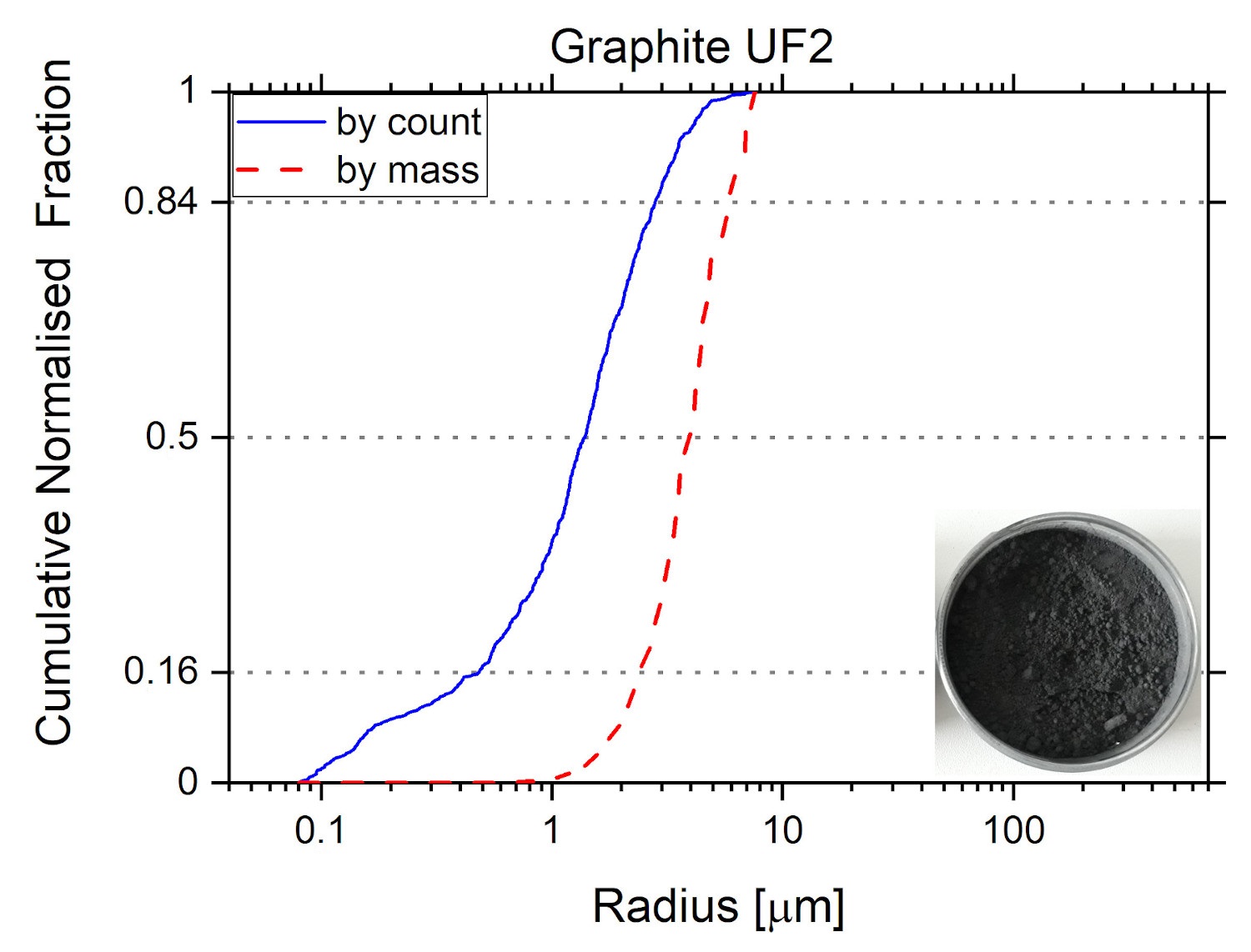}
    \end{minipage}
    \caption{Left: Microscopic caption of Graphite UF2. Right: Size distributions of Graphite UF2 calculated by count and by mass. The median and the $1 \sigma$ interval is indicated by the dotted lines. A macroscopic picture of the sample material is also shown.}
    \label{fig:GV_GraphiteUF2}
\end{figure*}

\begin{figure*}
    \centering
    \begin{minipage}[b]{0.45\textwidth}
    \includegraphics[width=\columnwidth]{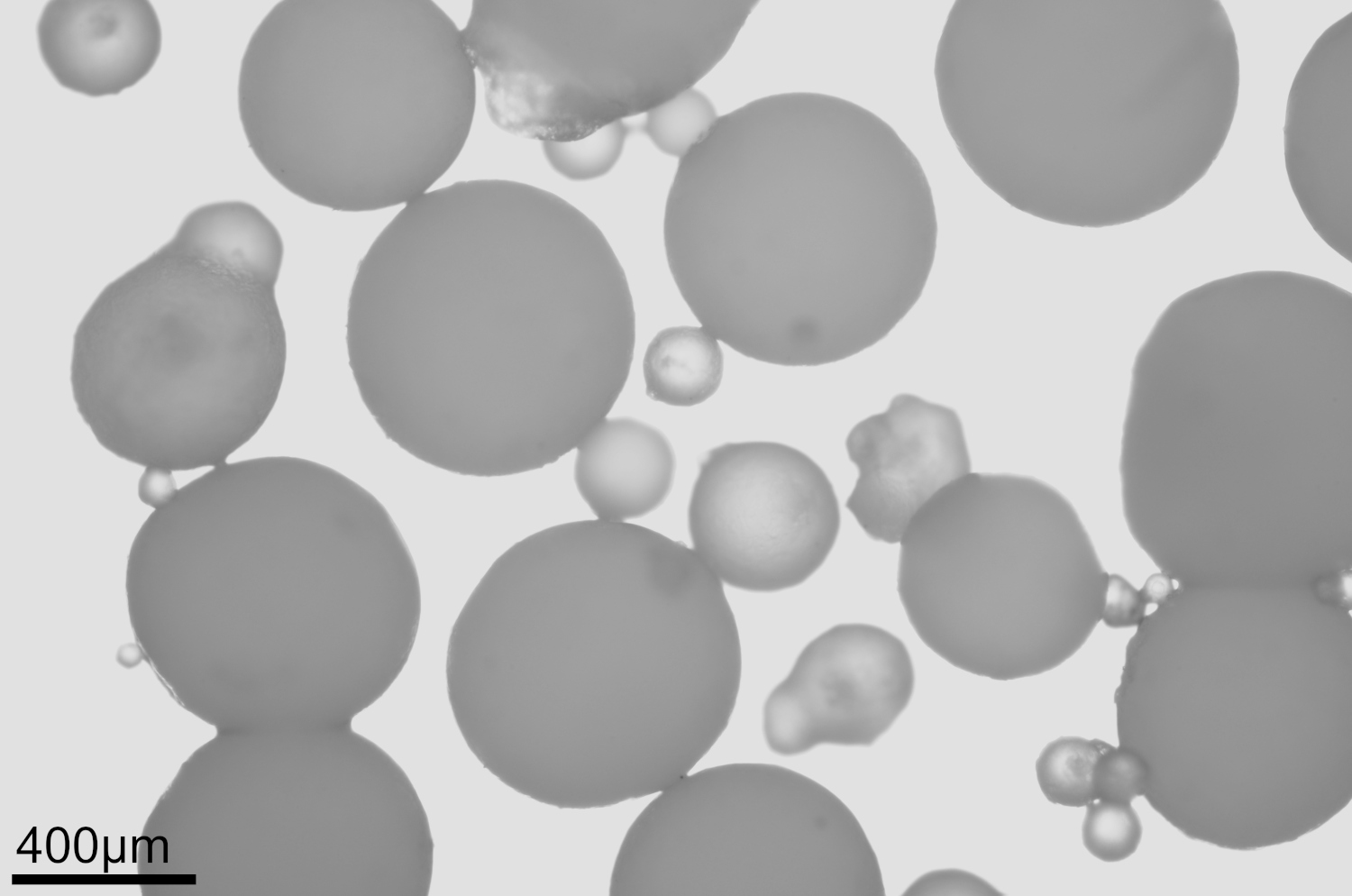}
    \end{minipage}
    \begin{minipage}[b]{.45\textwidth}
    \includegraphics[width=\columnwidth]{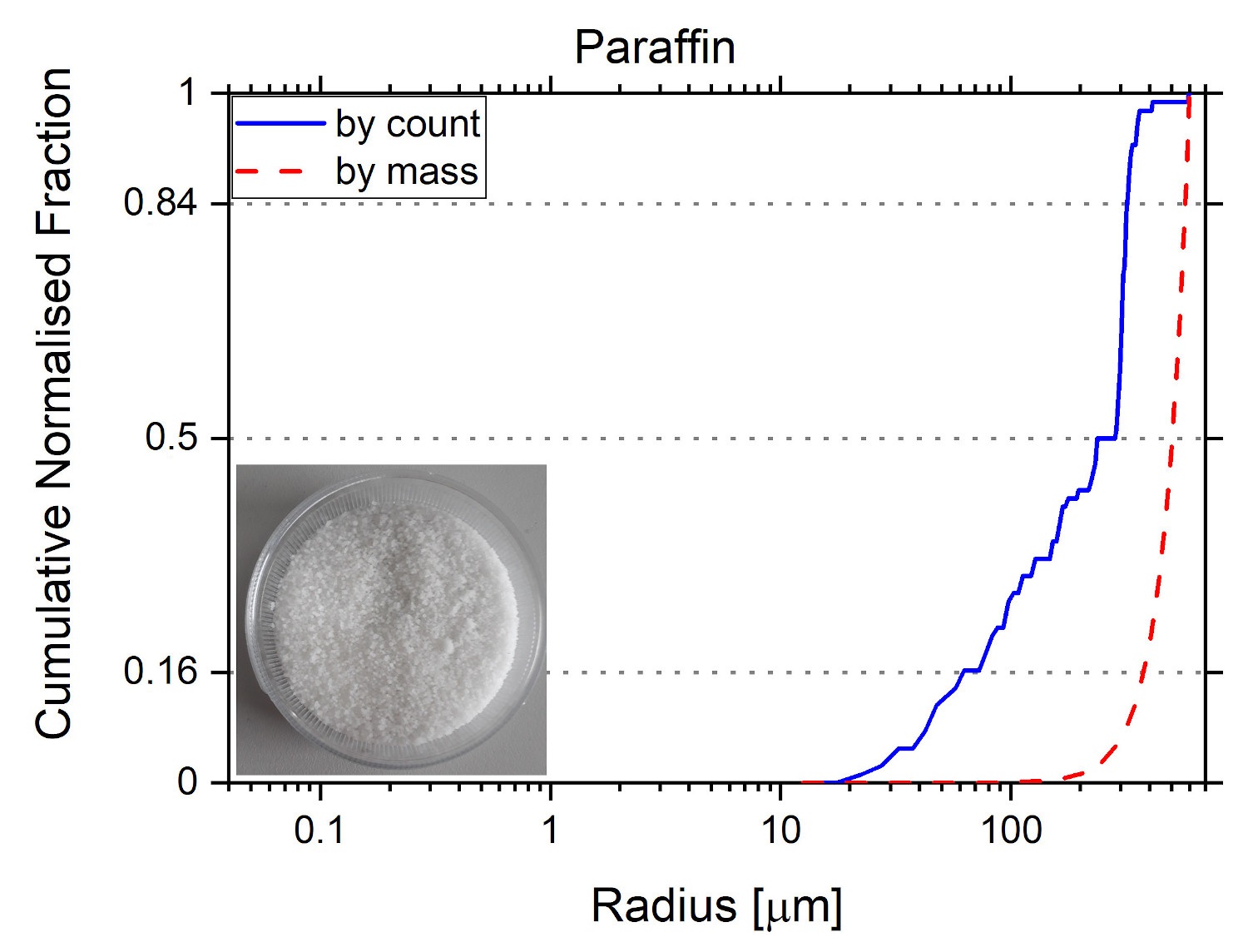}
    \end{minipage}
    \caption{Left: Microscopic caption of paraffin. Right: Size distributions of paraffin calculated by count and by mass. The median and the $1 \sigma$ interval is indicated by the dotted lines. A macroscopic picture of the sample material is also shown.}
    \label{fig:GV_Paraffin}
\end{figure*}

\begin{figure*}
    \centering
    \begin{minipage}[b]{0.45\textwidth}
    \includegraphics[width=\columnwidth]{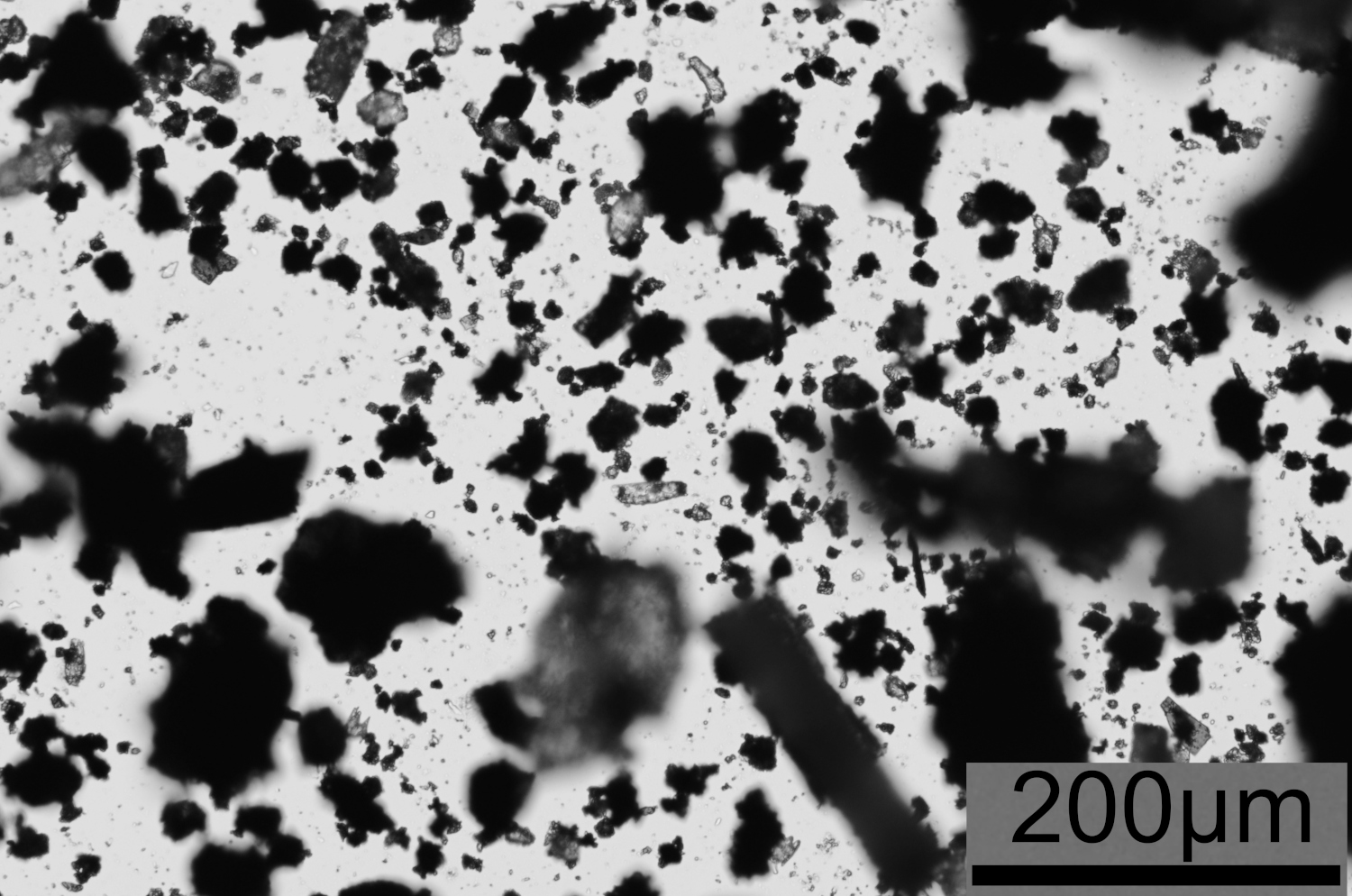}
    \end{minipage}
    \begin{minipage}[b]{.45\textwidth}
    \includegraphics[width=\columnwidth]{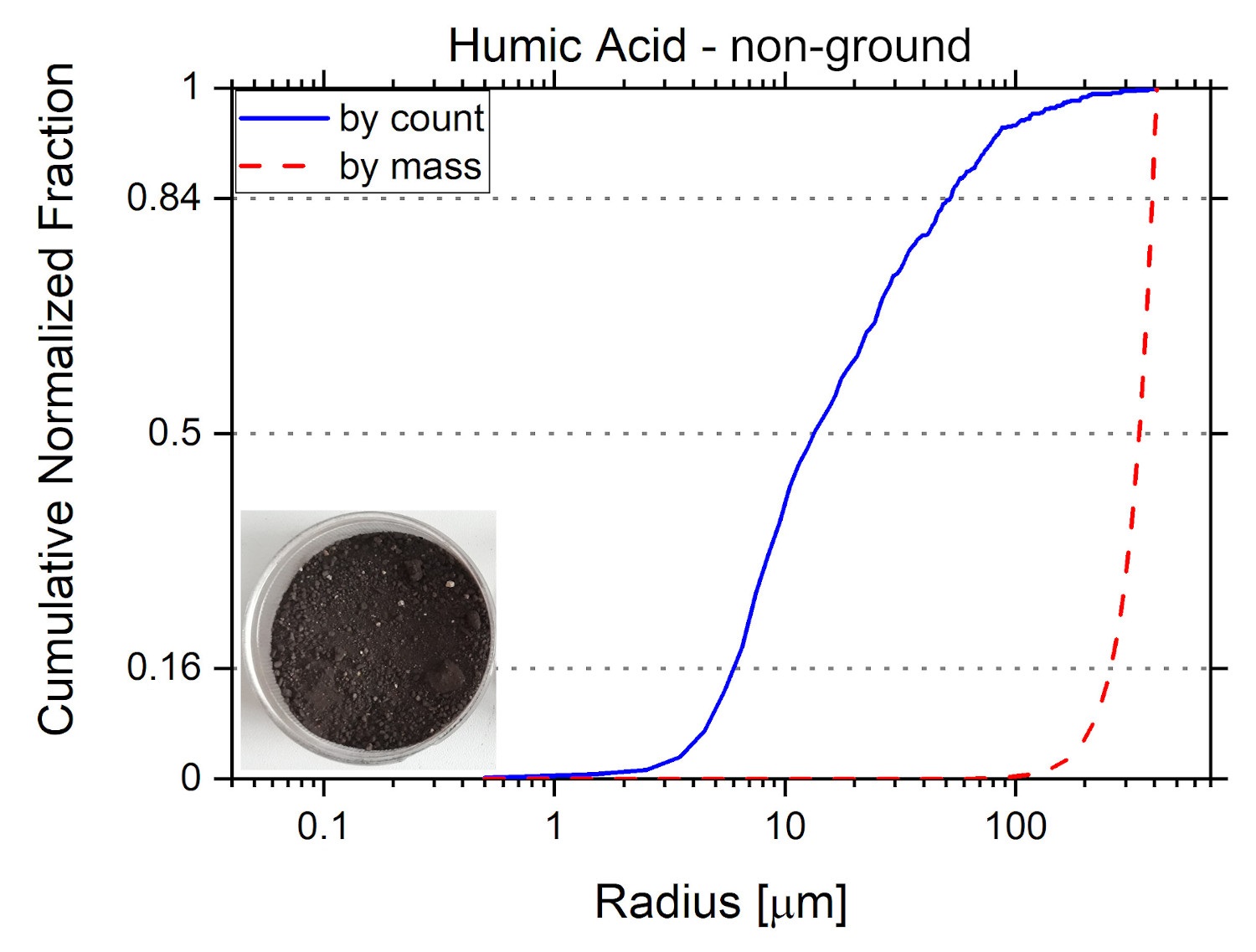}
    \end{minipage}
    \caption{Left: Microscopic caption of non-ground humic acids. Right: Size distributions of non-ground humic acid calculated by count and by mass. The median and the $1 \sigma$ interval is indicated by the dotted lines. A macroscopic picture of the sample material is also shown.}
    \label{fig:GV-Huminsaure-ungem}
\end{figure*}

\begin{figure*}
    \centering
    \begin{minipage}[b]{0.45\textwidth}
    \includegraphics[width=\columnwidth]{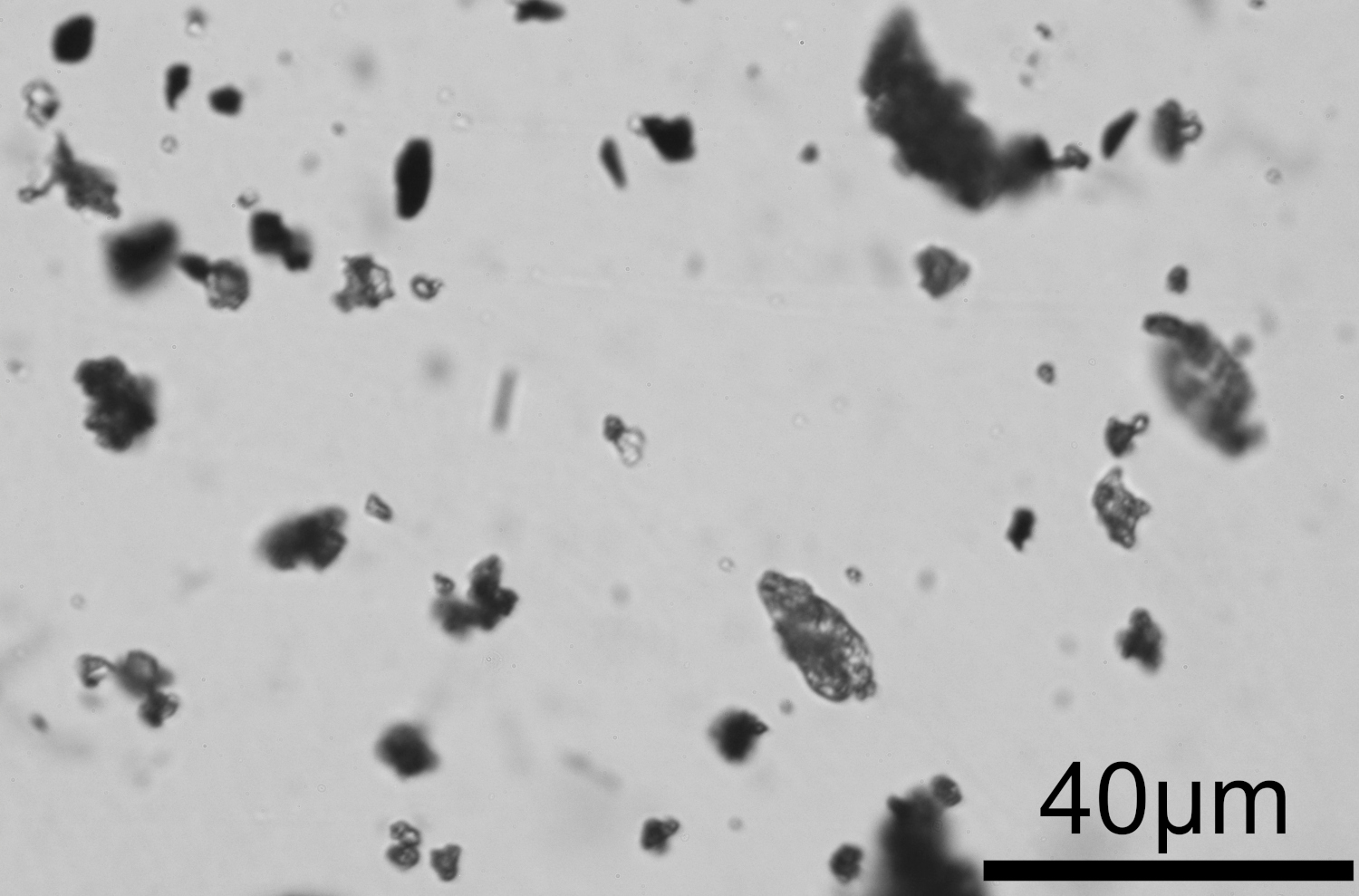}
    \end{minipage}
    \begin{minipage}[b]{.45\textwidth}
    \includegraphics[width=\columnwidth]{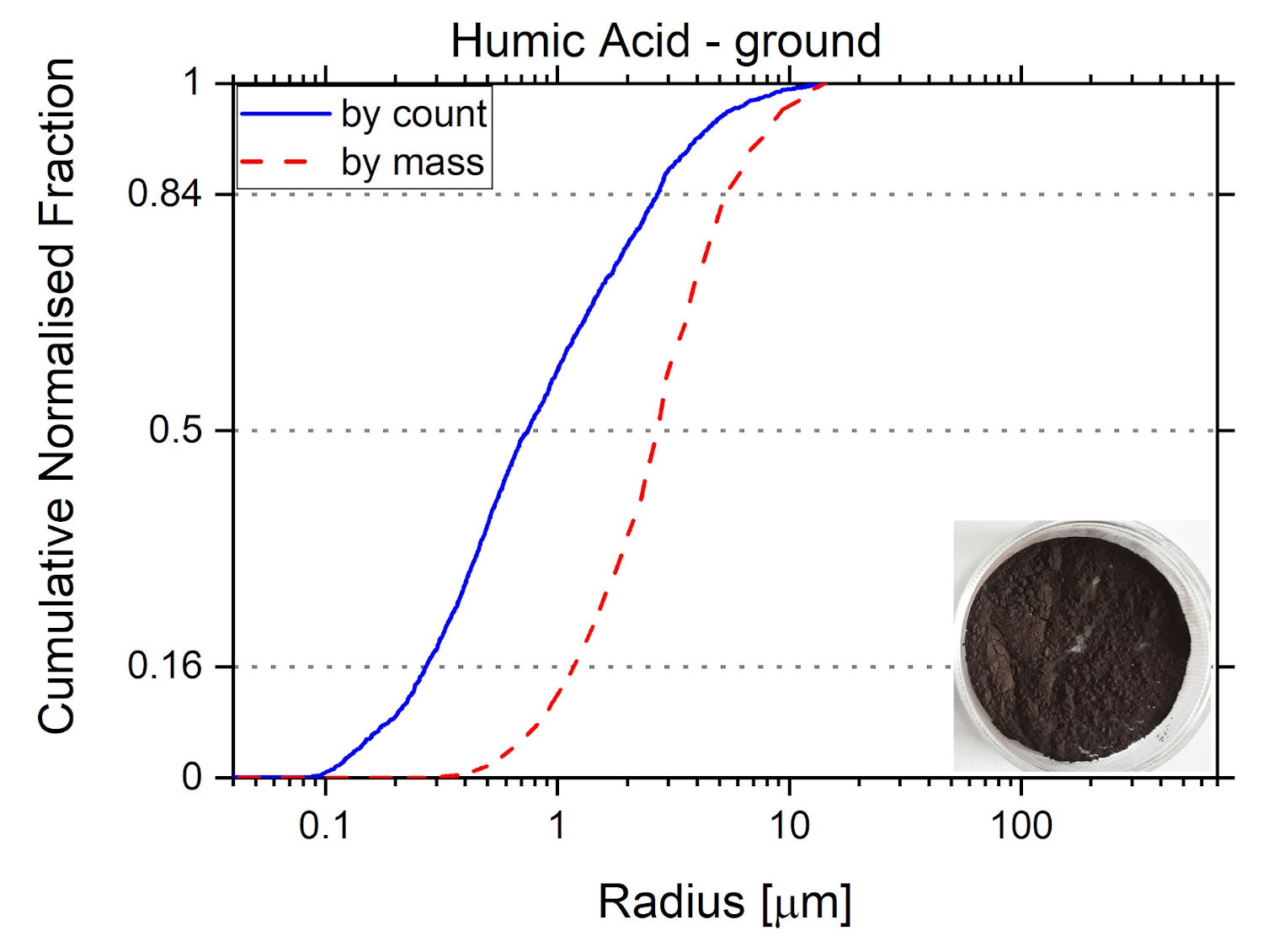}
    \end{minipage}
    \caption{Left: Microscopic caption of ground humic acids. Right: Size distributions of ground humic acid calculated by count and by mass. The median and the $1 \sigma$ interval is indicated by the dotted lines. A macroscopic picture of the sample material is also shown.}
    \label{fig:GV-Huminsaure-gem}
\end{figure*}

\begin{figure*}
    \centering
    \begin{minipage}[b]{0.45\textwidth}
    \includegraphics[width=\columnwidth]{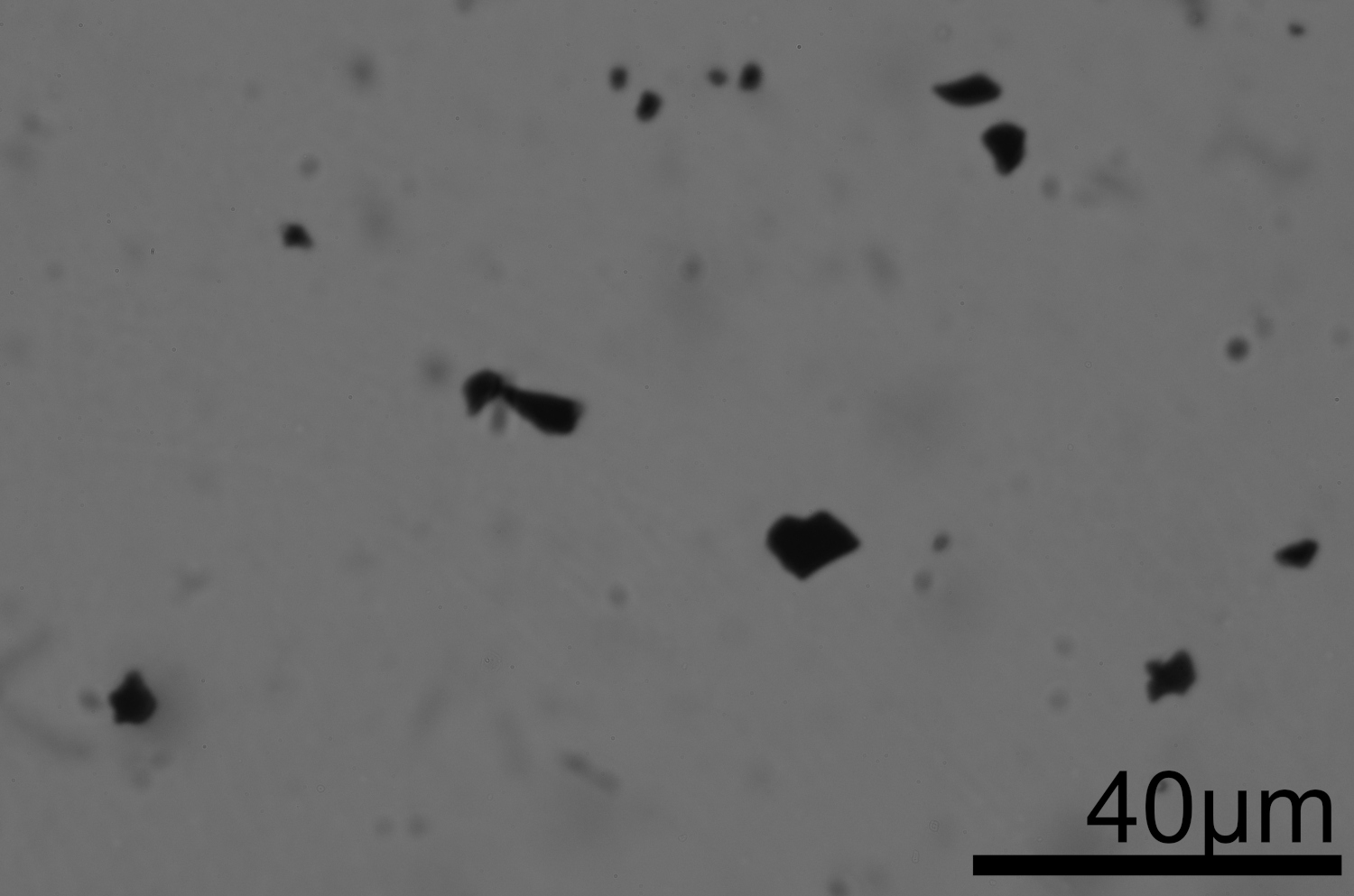}
    \end{minipage}
    \begin{minipage}[b]{.45\textwidth}
    \includegraphics[width=\columnwidth]{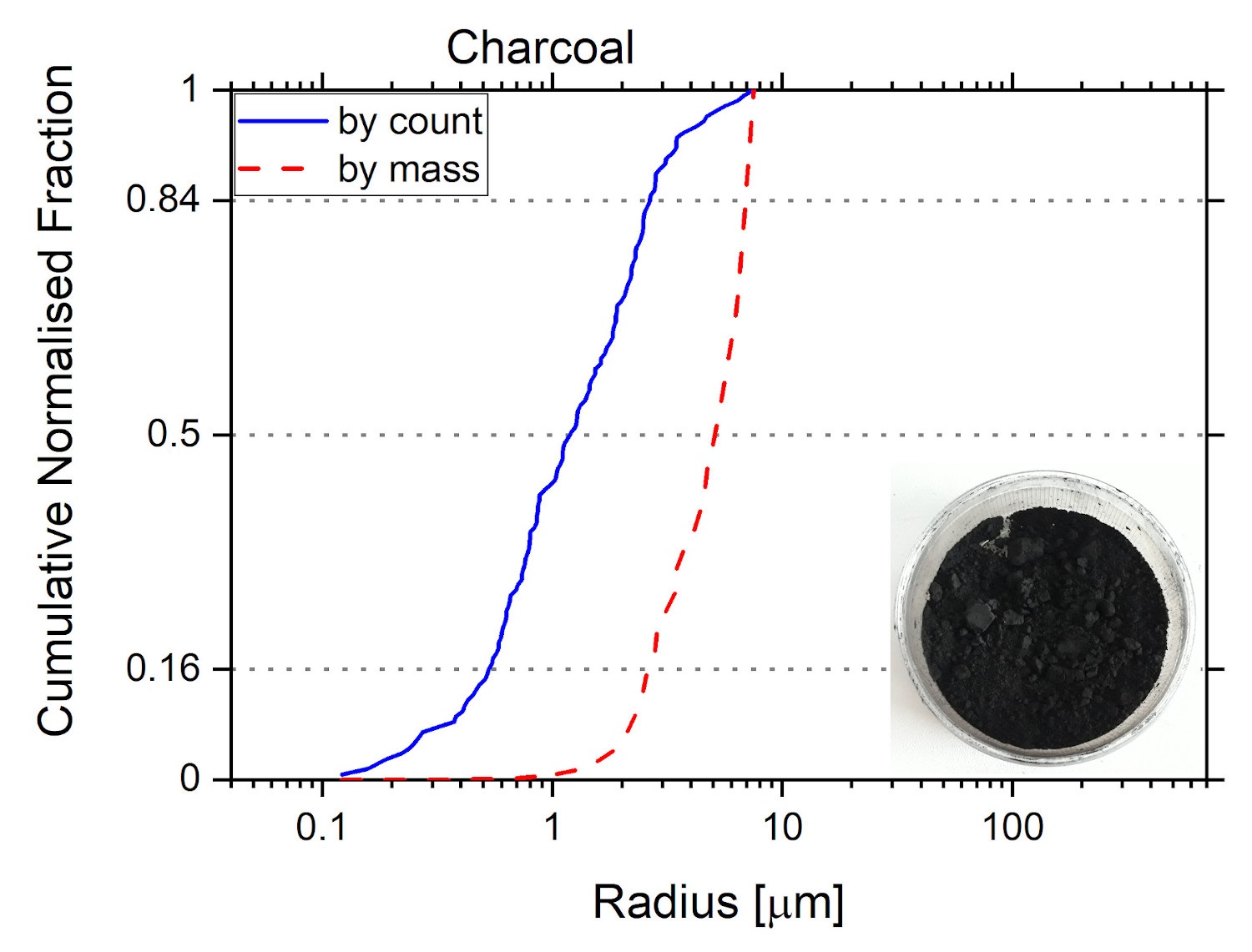}
    \end{minipage}
    \caption{Left: Microscopic caption of charcoal. Right: Size distributions of charcoal calculated by count and by mass. The median and the $1 \sigma$ interval is indicated by the dotted lines. A macroscopic picture of the sample material is also shown.}
    \label{fig:GV_Charcoal}
\end{figure*}

\begin{figure*}
    \centering
    \begin{minipage}[b]{0.45\textwidth}
    \includegraphics[width=\columnwidth]{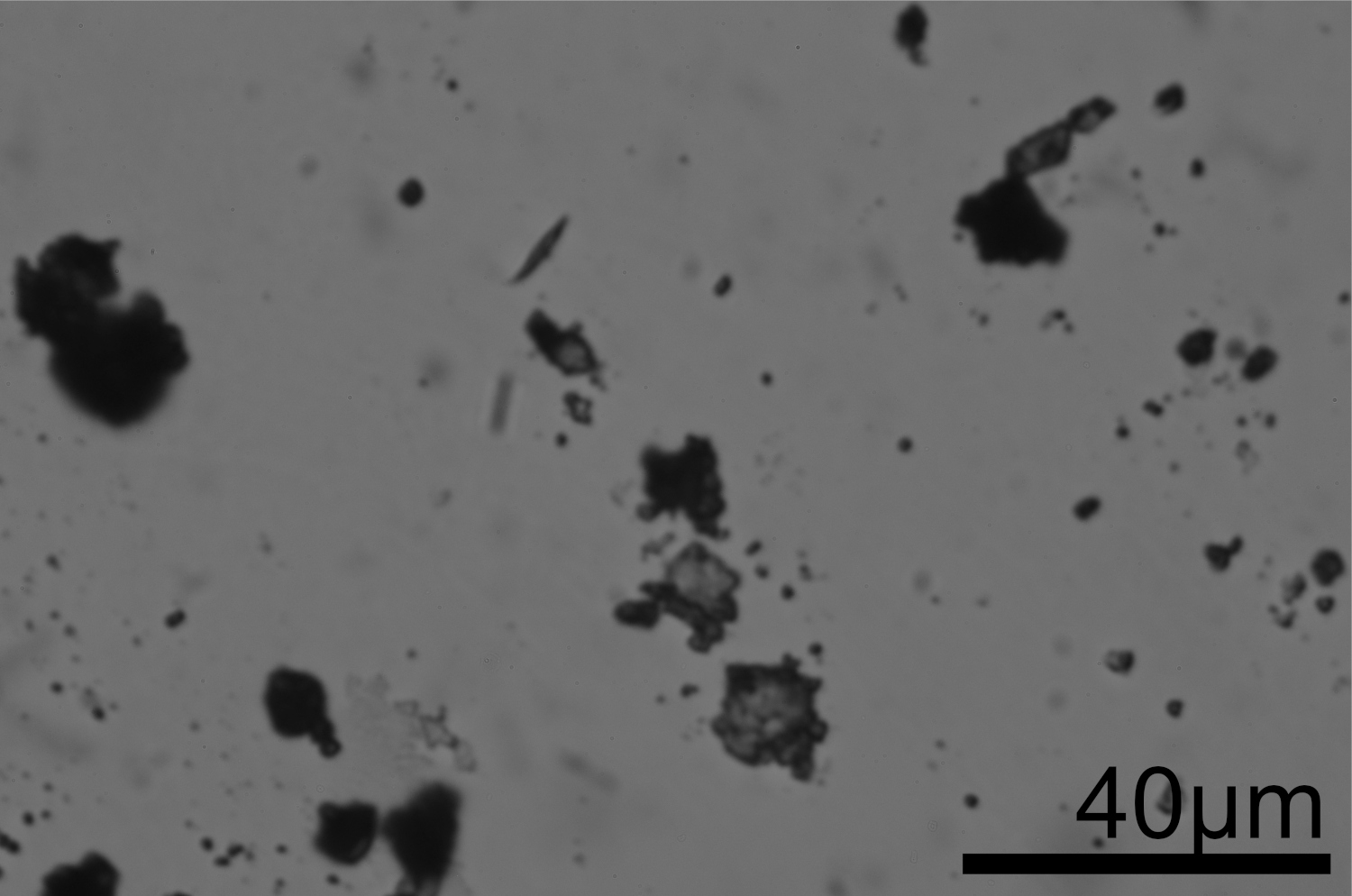}
    \end{minipage}
    \begin{minipage}[b]{.45\textwidth}
    \includegraphics[width=\columnwidth]{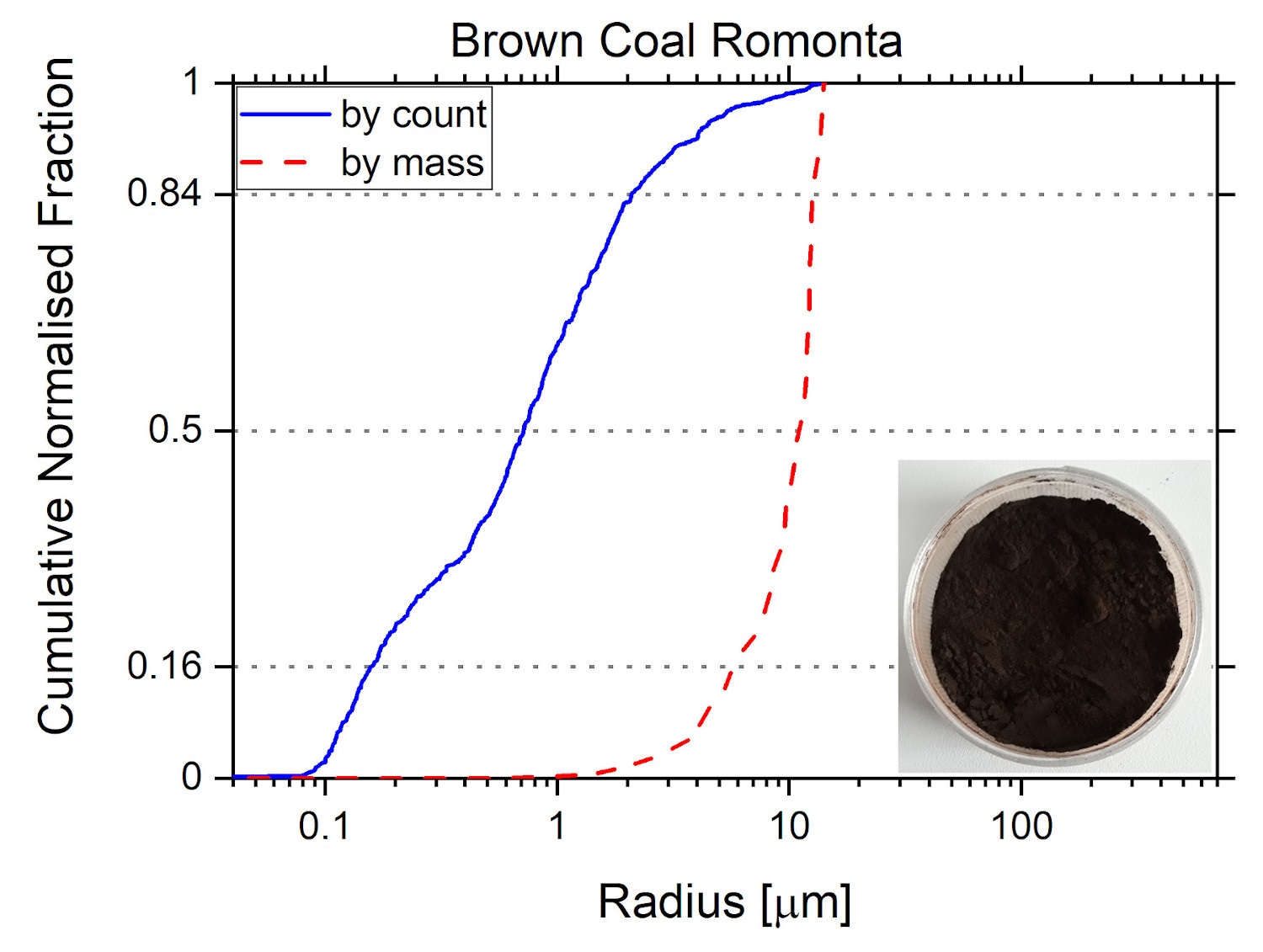}
    \end{minipage}
    \caption{Left: Microscopic caption of brown coal from Romonta. Right: Size distributions of brown coal from Romonta calculated by count and by mass. The median and the $1 \sigma$ interval is indicated by the dotted lines. A macroscopic picture of the sample material is also shown.}
    \label{fig:GV_BK_Romonta}
\end{figure*}

\begin{figure*}
    \centering
    \begin{minipage}[b]{0.45\textwidth}
    \includegraphics[width=\columnwidth]{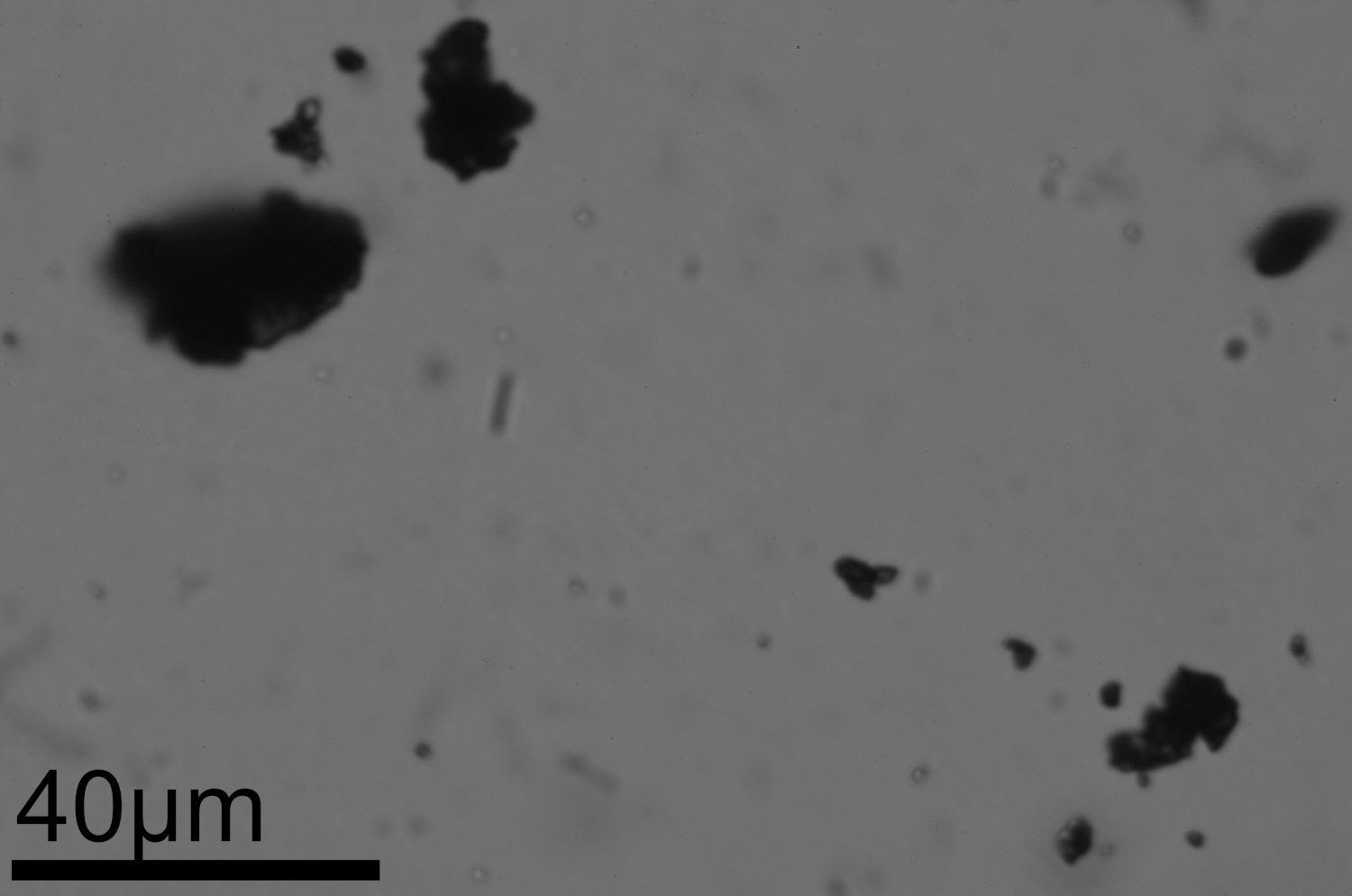}
    \end{minipage}
    \begin{minipage}[b]{.45\textwidth}
    \includegraphics[width=\columnwidth]{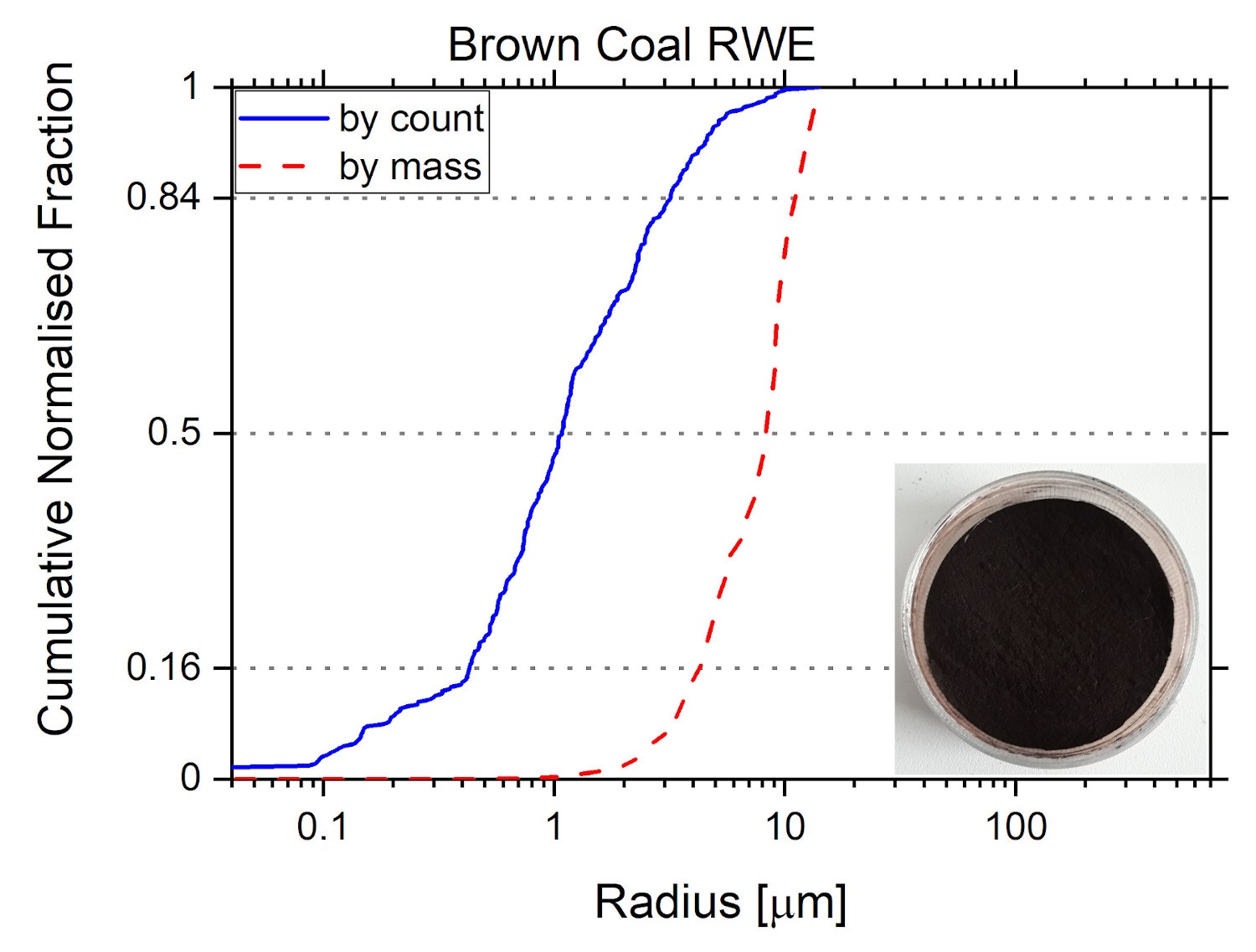}
    \end{minipage}
    \caption{Left: Microscopic caption of brown coal from RWE. Right: Size distributions of brown coal from RWE calculated by count and by mass. The median and the $1 \sigma$ interval is indicated by the dotted lines. A macroscopic picture of the sample material is also shown.}
    \label{fig:GV_BK_RWE}
\end{figure*}

\begin{figure*}
    \centering
    \begin{minipage}[b]{0.42\textwidth}
    \includegraphics[width=\columnwidth]{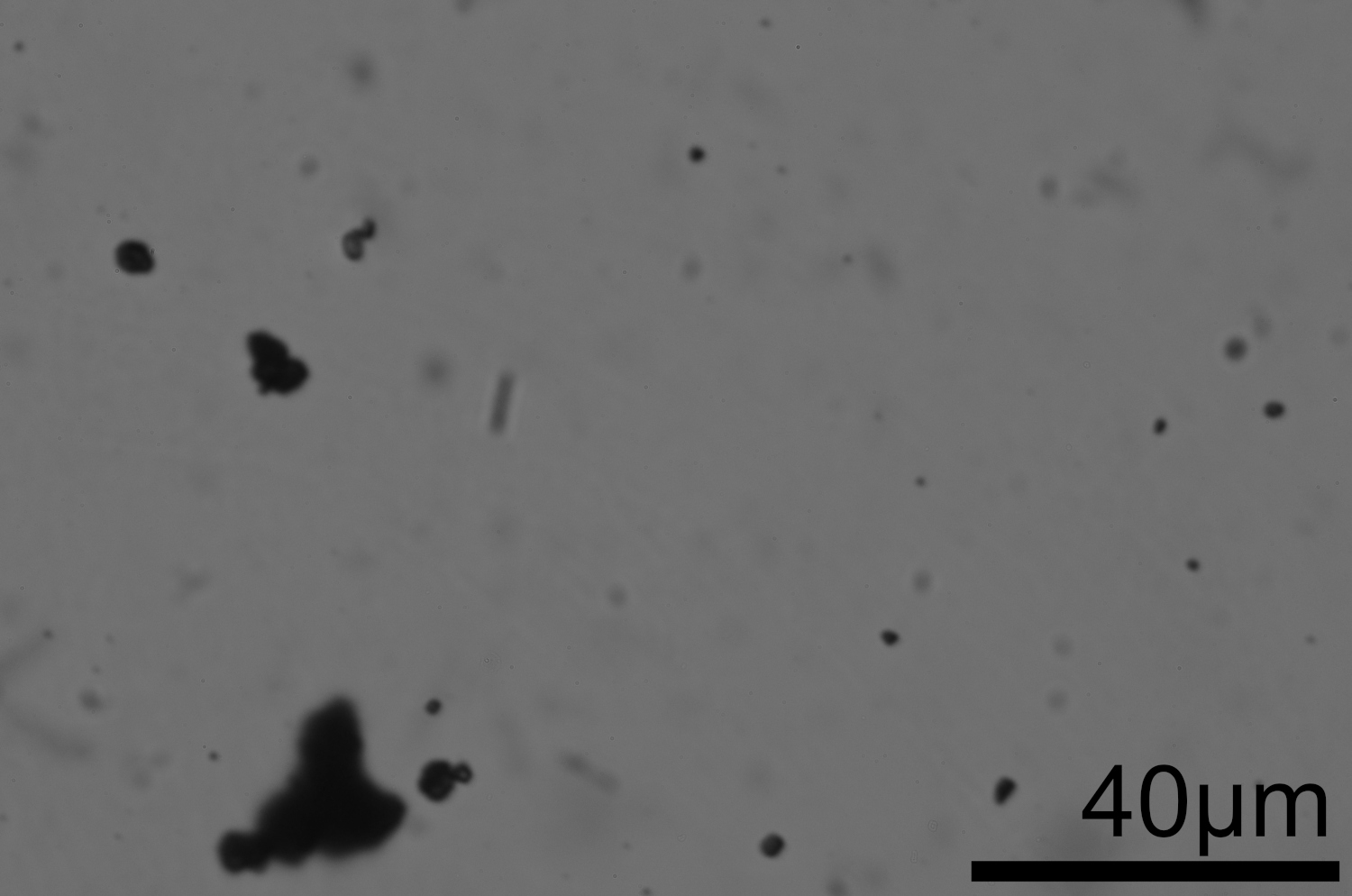}
    \end{minipage}
    \begin{minipage}[b]{.42\textwidth}
    \includegraphics[width=\columnwidth]{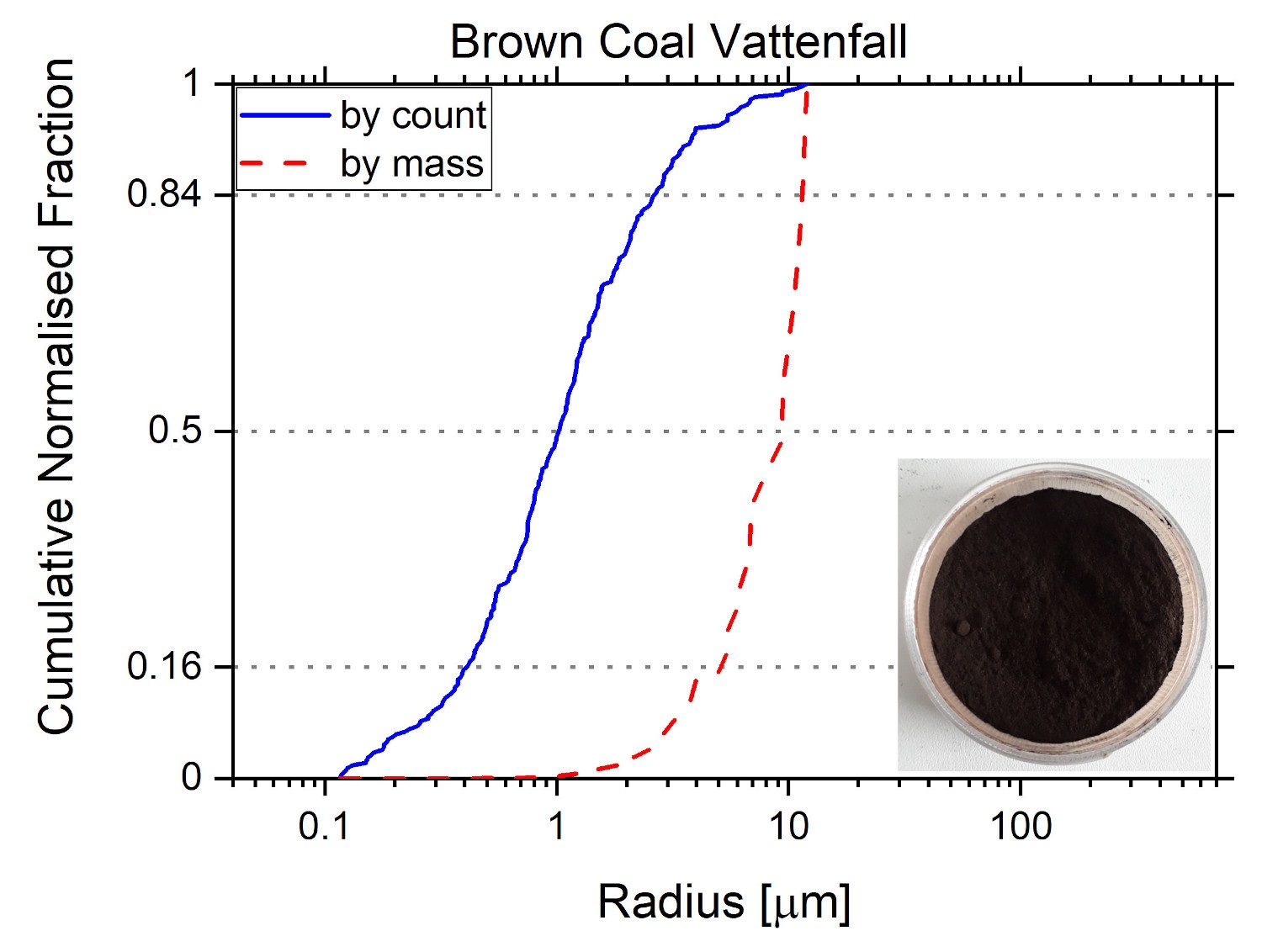}
    \end{minipage}
    \caption{Left: Microscopic caption of brown coal from Vattenfall. Right: Size distributions of brown coal from Vattenfall calculated by count and by mass. The median and the $1 \sigma$ interval is indicated by the dotted lines. A macroscopic picture of the sample material is also shown.}
    \label{fig:GV_BK_Vattenfall}
\end{figure*}

\begin{figure*}
    \centering
    \begin{minipage}[b]{0.42\textwidth}
    \includegraphics[width=\columnwidth]{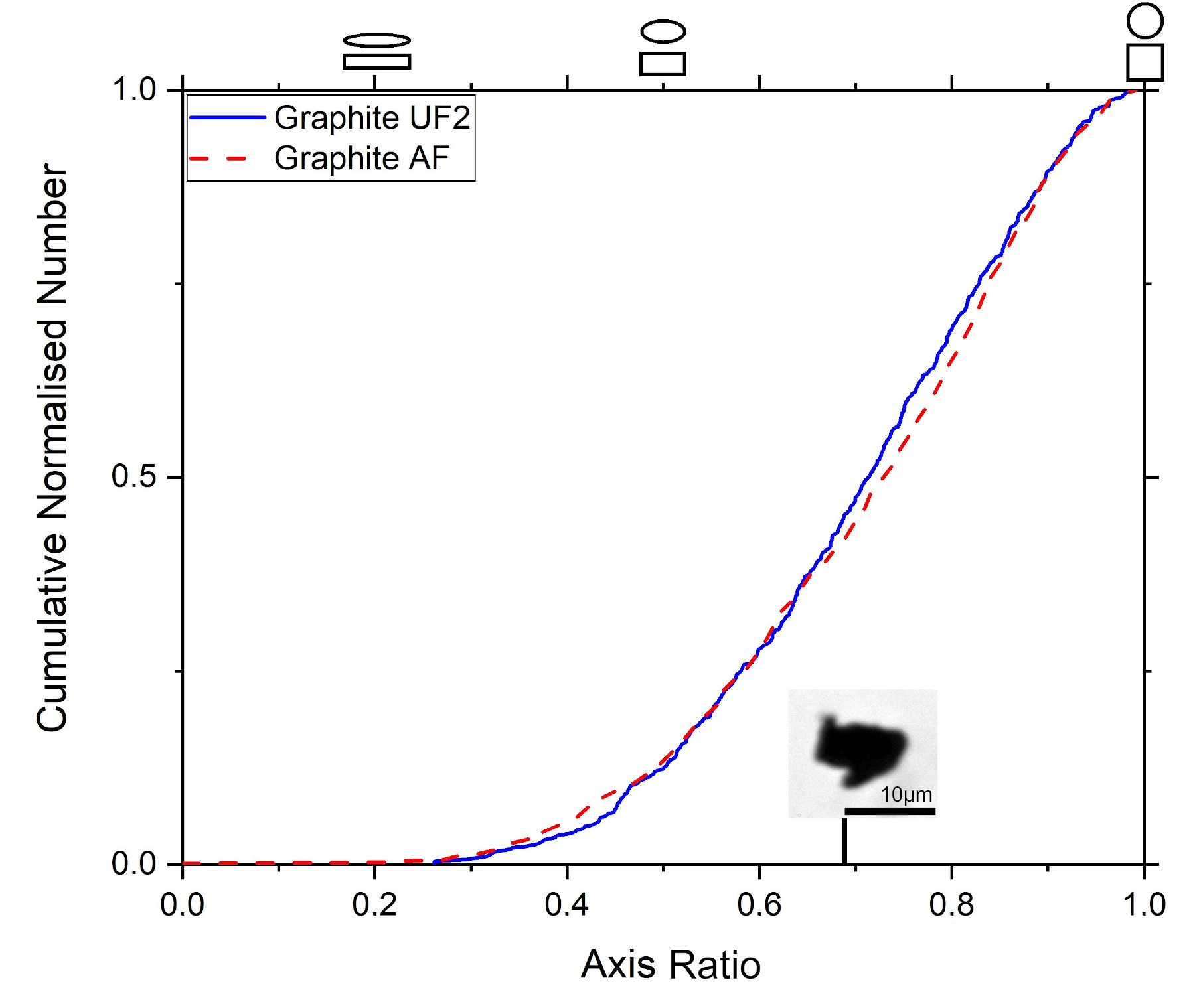}
    \end{minipage}
    \begin{minipage}[b]{0.42\textwidth}
    \includegraphics[width=\columnwidth]{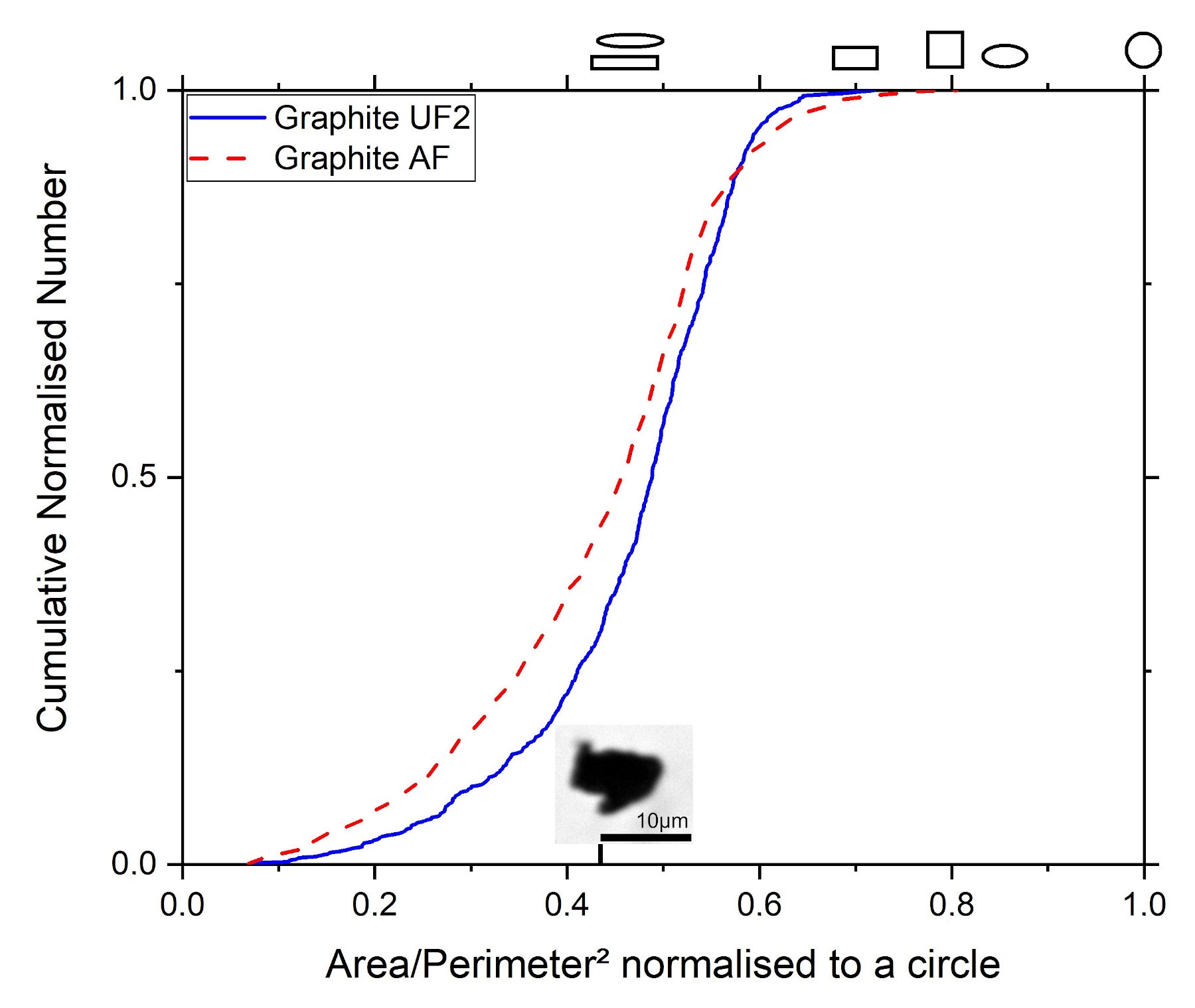}
 \end{minipage}
     \caption{Left: Axis ratio $C$ of Graphite AF and UF2. The axis ratio is defined as the ratio of semi-minor to semi-major axis. Right: Parameter $B$ of Graphite AF and UF2. Exemplarily a circle, two ellipses, a square, two rectangles and a particle are shown at their associated values. }
    \label{fig:form_graphite}
\end{figure*}

\begin{figure*}
    \centering
    \begin{minipage}[b]{0.42\textwidth}
    \includegraphics[width=\columnwidth]{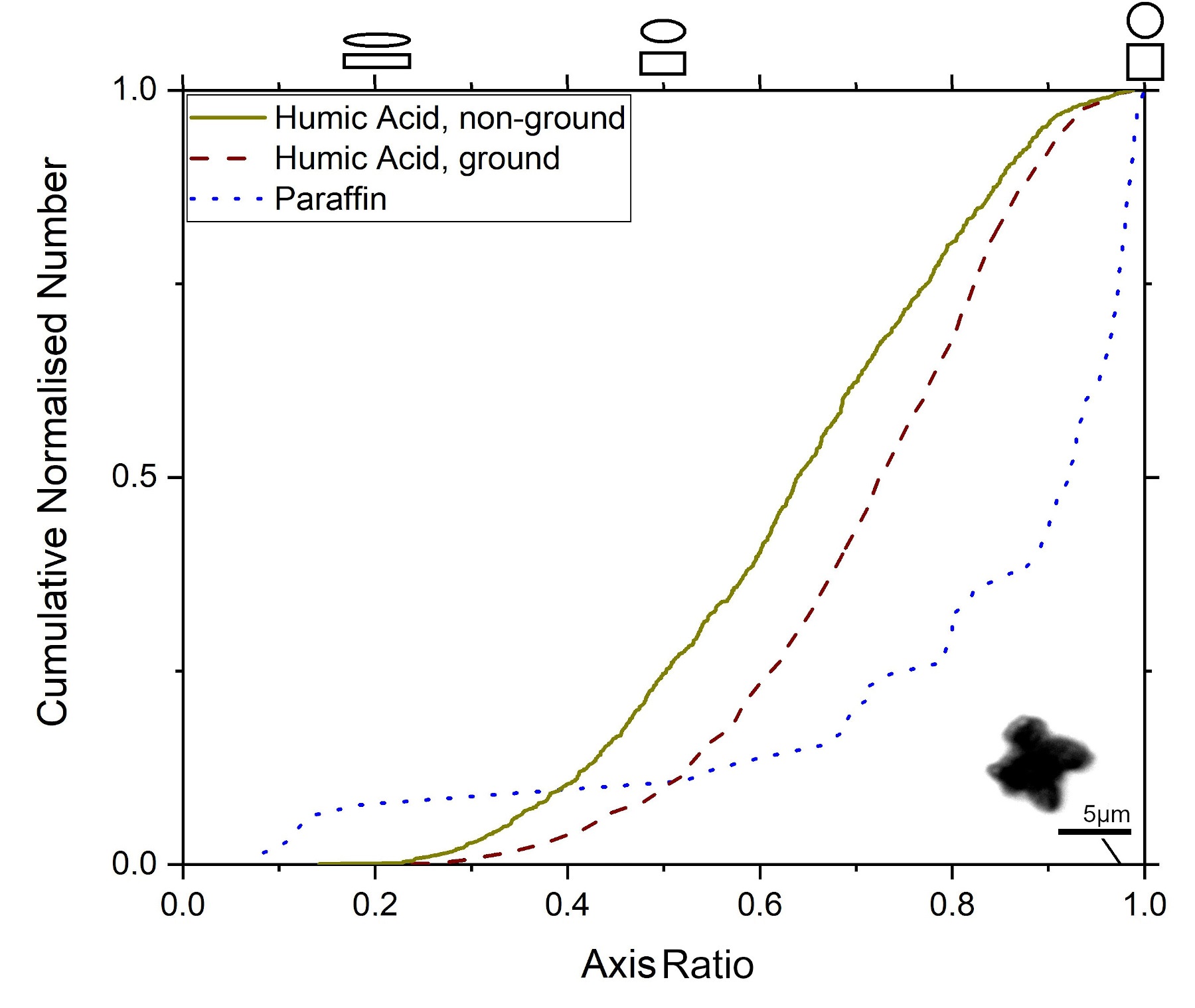}
    \end{minipage}
    \begin{minipage}[b]{0.42\textwidth}
    \includegraphics[width=\columnwidth]{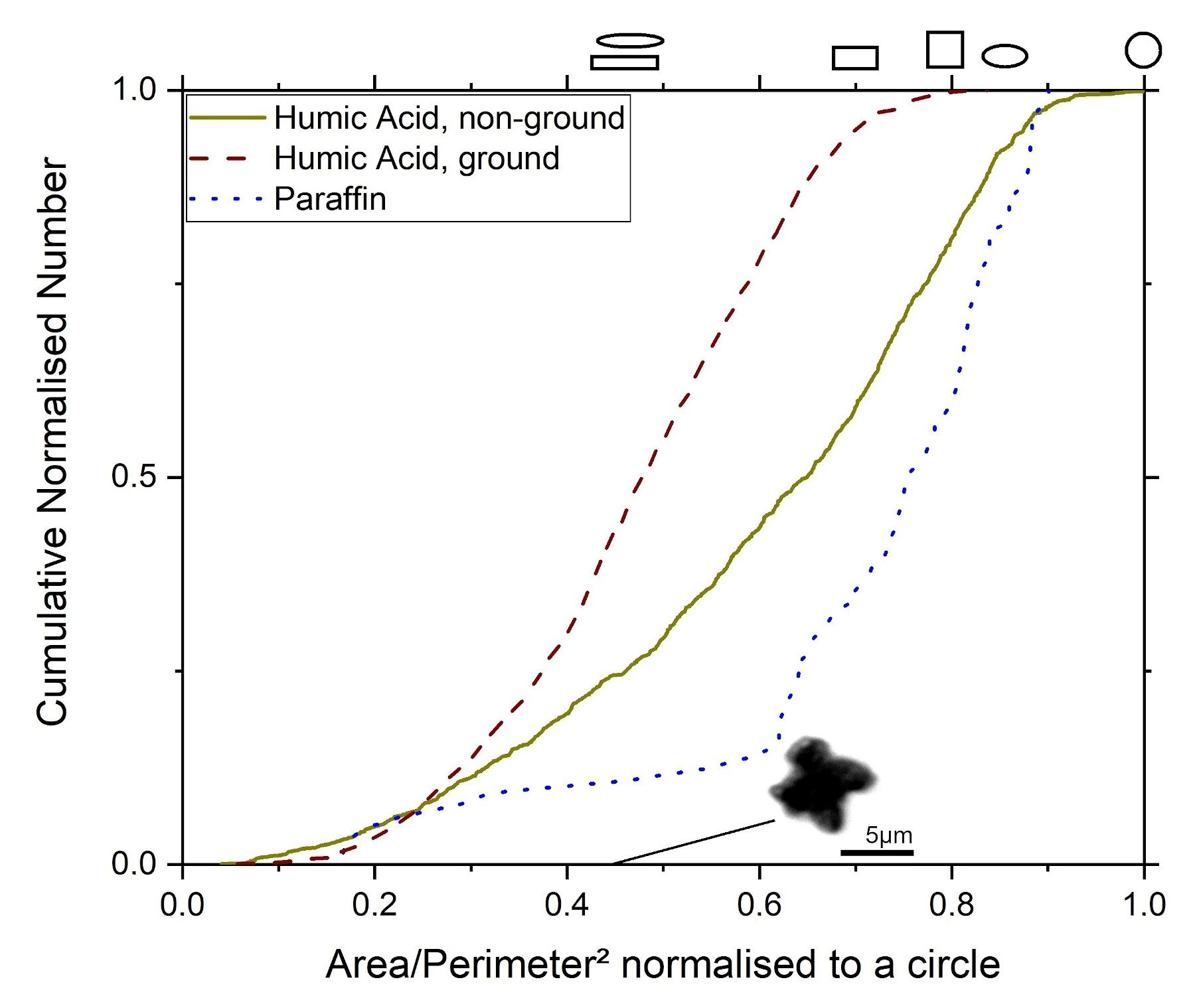}
    \end{minipage}
    \caption{Left: Axis ratio $C$ of humic acids and paraffin. The axis ratio is defined as the ratio of semi-minor to semi-major axis. Right: Parameter $B$ of humic acids and paraffin. Exemplarily a circle, two ellipses, a square, two rectangles and a particle are shown at their associated values.}
    \label{fig:Form_H_P}
\end{figure*}

\begin{figure*}
    \centering
    \begin{minipage}[b]{0.43\textwidth}
    \includegraphics[width=\columnwidth]{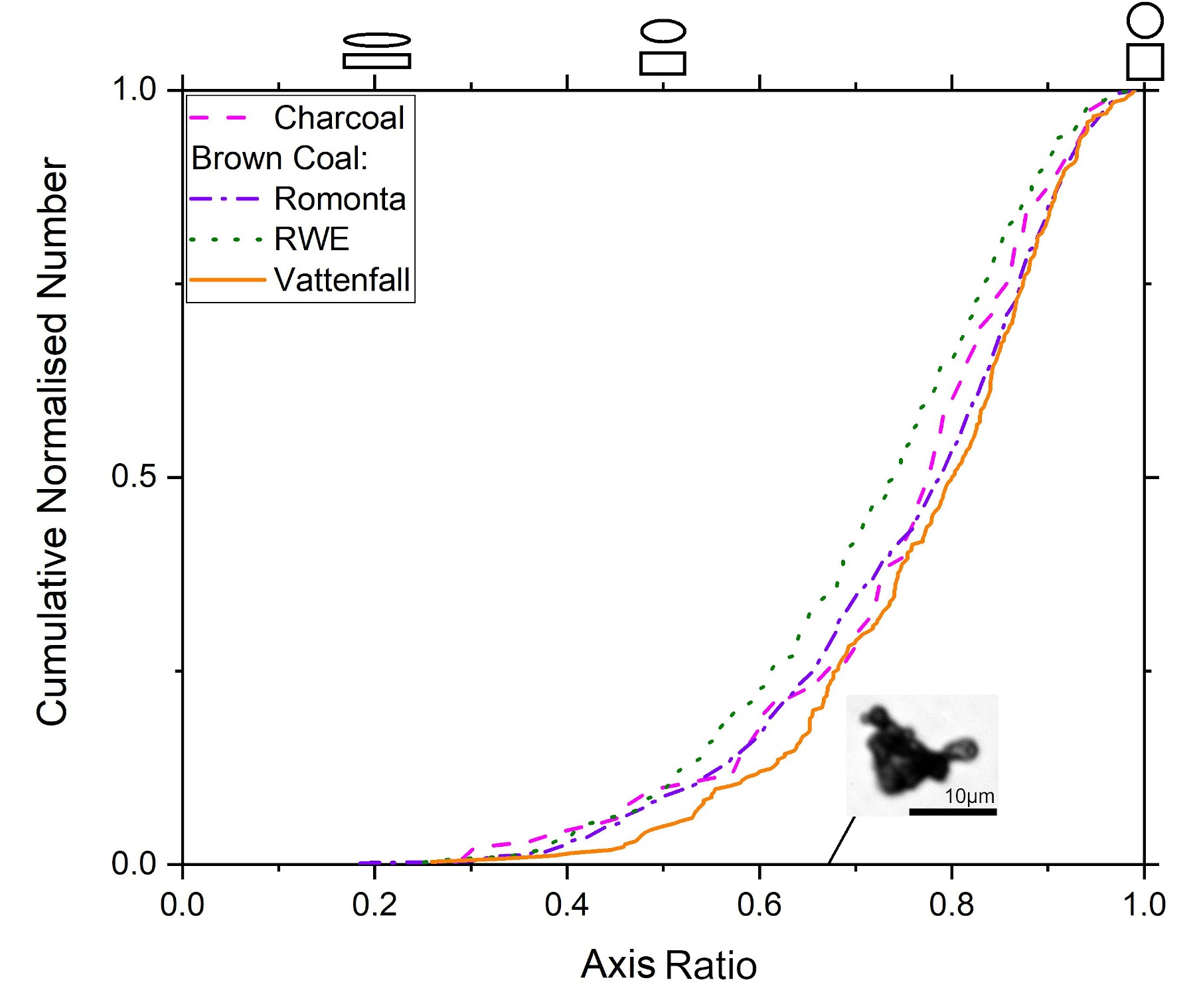}
    \end{minipage}
    \begin{minipage}[b]{0.43\textwidth}
    \includegraphics[width=\columnwidth]{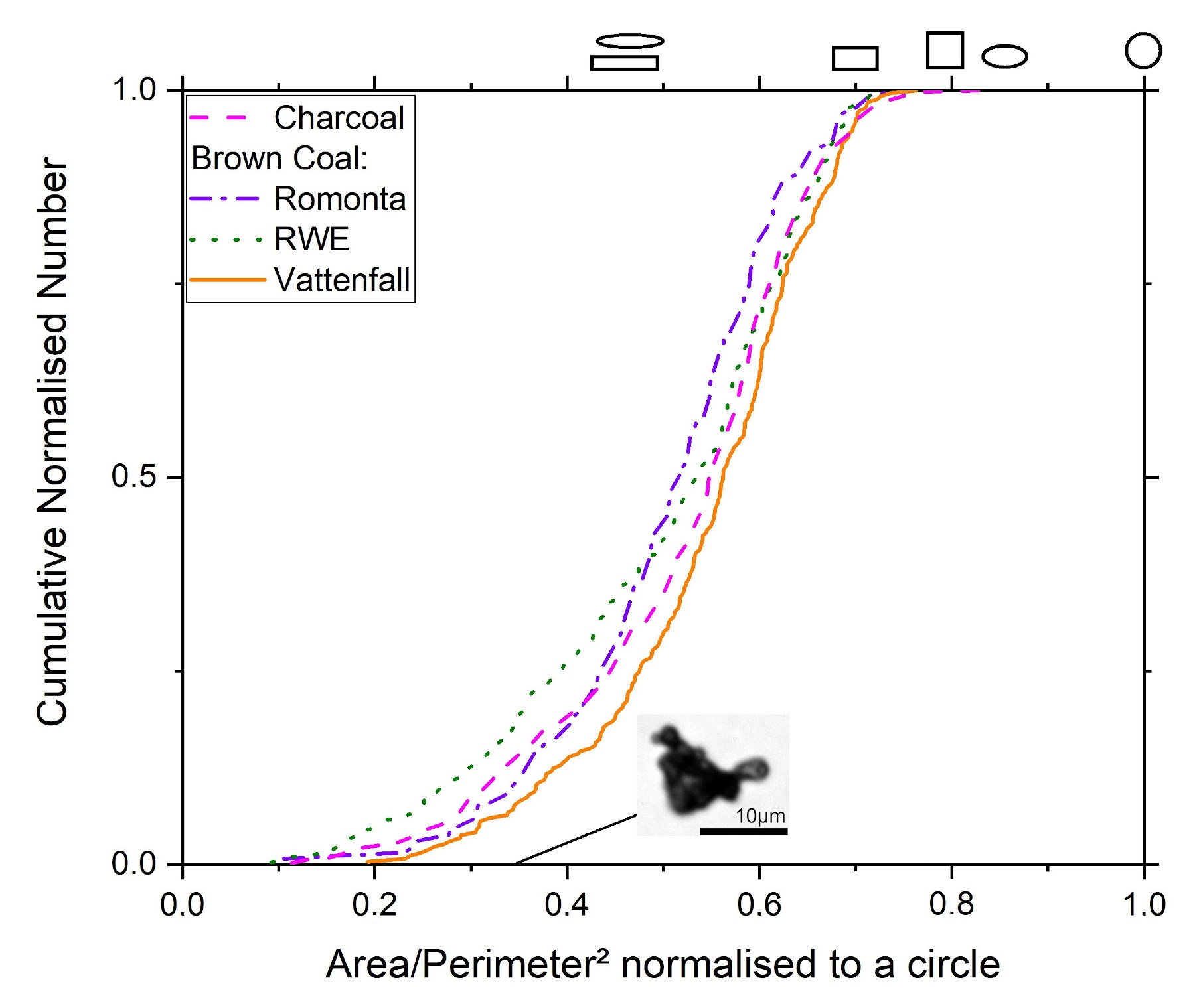}
    \end{minipage}
    \caption{Left: Axis ratio $C$ of the brown-coals and the charcoal.The axis ratio is defined as the ratio of semi-minor to semi-major axis. Right: Parameter $B$ of the brown-coals and the charcoal. Exemplarily a circle, two ellipses, a square, two rectangles and a particle are shown at their associated values.}
    \label{fig:Form_coals}
\end{figure*}

With the help of microscopic pictures, the grain size distributions of the used micro-granular materials were determined. Therefore, the material was dispersed in water or ethanol and spread on an object slide to separate the individual grains. Several microscope pictures were captured and analysed for each material. An ellipse was fitted to the particles and the semi-major and semi-minor axes were measured. To describe the shape of the particles, the axis ratio $C$ was defined as the ratio of the semi-minor and semi-major axis:
\begin{equation}
  C = \frac{\textrm{semi-major\,axis}}{\textrm{semi-minor\,axis}}.  
\end{equation} A value near $C=1$ means that the projected particle shape is close to circular, whereas a smaller value describes more elongated shapes. For considering also edges in the shape, another parameter $B$ was defined, which describes the ratio of the cross section $A$ to square of the perimeter $U$ of the particle, normalised to that of a circle. Here, we used real cross section and perimeter values of the particles and not the fitted ellipses. Thus, the equation
\begin{equation}
    B = \frac{4 \pi A}{U^2}
\end{equation}
defines this parameter $B$. The ratio of the square of the perimeter to the cross section of a circle equals 4$\pi$. Due to the normalisation to a circle, a value close to $B=1$ means a near-spherical shape and smaller values describe a more complex shape. To illustrate the used parameters of particle shape, examples are given in the following plots. The associated values of a circle, two ellipses, a square, two rectangles and an exemplary particle are marked, respectively. In addition, for each material, a microscopic caption is presented. This caption and the size distribution, as well as macroscopic pictures for each material can be seen in Figures \ref{fig:GV_GraphiteAF} to \ref{fig:GV_BK_Vattenfall}. The axis ratio values $C$ and the values of parameter $B$ are shown in Figures \ref{fig:form_graphite} to \ref{fig:Form_coals}. 

On the basis of the shape parameters, the different materials can be compared and similarities as well as differences can be noted. The graphite powders seem to be very similar in shape as their curves of axis ratio nearly cover each other. Graphite AF shows smaller values in $B$ than Graphite UF2. This can be interpreted as the particles of AF being more angled. In both cases, the axis ratio is larger than the parameter $B$, which means that the particles are more angled than elongated. The ground humic acid shows a similar shape as the graphites, however, the differences between ground and non-ground humic acid are larger. Whereas the non-ground material features a smaller axis ratio, the ground material shows smaller $B$ values. This means that the non-ground humic acid consists of smoother, less compact particles and the particles of the ground humic acid are more compact and more angled. This could be due to the large difference in grain size. 
The grains of paraffin are the roundest ones. Particles, differing from spheres, are more elliptical than angled. The coals are very similar to the graphites. They do not show very spherical grains, but angled ones and it seems like they are very complex in shape. Among each other, the coals differ rarely. 



\newpage
\section{Parameter of the Volume Filling Factor Fit}
\label{appendix:vff_fit_parameter}
\begin{table}
\centering
\caption{Fitting parameter $a$ and $b$ for the fit of the tensile strength as a function of the volume filling factor, see Equation \ref{eq:VFF_Fit}, as well as the $R^2$-value for the fit quality and the tensile strength for a volume filling factor of $\Phi=0.5$. For humic acid and the coals the highest and lowest $a$ value were chosen and fixed for the fit.}
\label{tab:fit_parameter}
\begin{tabular}{|c|c|c|c|c|}
\hline
Material  & Parameter $a$    & Parameter $b$ & $R^2$ & $\sigma(\Phi=0.5)$ \\ 
& & &  & $[\mathrm{kPa}]$ \\
\hline
Graphite  &  $6.2 \pm 1.3 $  & $0.26 \pm 0.05 $ & 0.91 & $32 \pm 6$ \\ 
(AF) & & & &\\
\hline
Graphite &   $6.3 \pm 0.6 $  &  $0.11 \pm 0.02$ & 0.98 & $280 \pm 40 $ \\ 
(UF2) & & & &\\
\hline
Humic Acid  &  $2.9$  &  $0.60 \pm 0.03 $ & 0.48 & $0.52 \pm 0.09$ \\ 
(non-ground)&  $6.3$  &  $0.61 \pm 0.02 $ & 0.59 & $0.19 \pm 0.05$\\
\hline
Humic Acid  &   $2.9$  &  $0.60 \pm 0.03$ & 0.56 & $0.52\pm0.08$ \\ 
(ground)    &   $6.3$  &  $0.66 \pm 0.02$ & 0.50 & $0.10\pm 0.01$ \\
\hline
Paraffin     & $2.9 \pm 0.6$  &  $0.39\pm0.07$ & 0.94 & $2.0\pm0.2$ \\ 
 &         &  &    &      \\
\hline
Brown Coal   &   $2.9$  &  $0.64 \pm 0.04$ & -0.47 & $0.39\pm0.04$  \\ 
Romonta      &   $6.3$  &  $0.60\pm0.02$ & -1.50 & $0.24\pm0.02$ \\
\hline
Brown Coal   &   $2.9$  &  $0.84 \pm 0.04$ & 0.35 & $0.10\pm0.02$  \\ 
RWE          &   $6.3$  &  $0.68\pm0.01$ & 0.60 & $0.08\pm0.01$ \\
\hline
Brown Coal   &   $2.9$  &  $0.82 \pm 0.02$ & 0.34 & $0.12\pm0.02$\\ 
Vattenfall   &   $6.3$  &  $0.66\pm0.01$ & 0.47 & $0.10\pm0.01$ \\
\hline
Charcoal     &   $2.9$  &  $0.63 \pm 0.03$ & 0.62 & $0.84\pm0.09$ \\ 
&                $6.3$  &  $0.55\pm0.01$ & 0.58 & $0.52\pm0.08$ \\
\hline
\end{tabular}
\end{table}

\label{lastpage}
\end{document}